\documentclass[journal, twocolumn]{IEEEtran}

\newcommand{\remove}[1]{}

\newtheorem{definition}{Definition}
\newtheorem{lemma}{Lemma}

\newtheorem{theorem}{Theorem}
\newtheorem{corollary}{Corollary}
\usepackage[cmex10]{amsmath}
\usepackage{amssymb}
\usepackage{mathrsfs}
\usepackage{graphicx}
\usepackage{color}
\usepackage{cite}
\usepackage{multirow}
\usepackage{framed}

\newcommand{\bI}{{\mathsf I}}
\newcommand{\bH}{{\mathsf H}}
\newcommand{\bP}{\mathsf{Pr}}

\newcommand{\cR}{{\cal R}}
\newcommand{\cC}{{\cal C}}
\newcommand{\cH}{{\cal H}}
\newcommand{\cL}{{\cal L}}
\newcommand{\cM}{{\cal M}}
\newcommand{\cX}{{\cal X}}
\newcommand{\cY}{{\cal Y}}
\newcommand{\cZ}{{\cal Z}}

\newcommand{\cS}{{\cal S}}

\newcommand{\bF}{{\mathbb F}}
\newcommand{\SD}{{\bf SD}}

\newcommand{\del}{{(0,\delta)}}

\newcommand{\enc}{\mathsf{AWTPenc}}
\newcommand{\dec}{\mathsf{AWTPdec}}

\newcommand{\cA}{{\cal A}}
\newcommand{\eps}{{\epsilon}}
\newcommand{\ADV}{{\bf Adv}}

\newcommand{\rorw}{{ (\rho_r, \rho_w) }}

\newcommand{\rd}{\color{red}}

\newcommand{\edwtp}{{$(\epsilon, \delta)$-AWTP}}
\newcommand{\ed}{{$(\epsilon, \delta)$}}

\newcommand{\pl}{\mathsf{Poly}}

\ifCLASSINFOpdf
\else
\fi

\begin{document} 

\title{A Model for Adversarial Wiretap Channel}

\author{Pengwei~Wang\IEEEauthorrefmark{1} and Reihaneh~Safavi-Naini\IEEEauthorrefmark{1} \\
\thanks{\IEEEauthorrefmark{1} This work is in part supported by Alberta Innovates Technology Futures, in the Province of Alberta, Canada.} 
Department
of Computer Science, University of Calgary,
Canada\\
 e-mail: [pengwwan,  rei]@ucalgary.ca 
}
\remove{
\markboth{Journal of \LaTeX\ Class Files,~Vol.~11, No.~4, December~2012}
}

\maketitle

\begin{abstract}
In wiretap model of secure communication, the goal is to provide (asymptotic) perfect secrecy and reliability over a noisy channel that is 
eavesdropped by an adversary with unlimited computational power. 
This goal is achieved by taking advantage of the channel noise and without 
requiring a shared key. The model has attracted considerable attention in recent years 
because it captures eavesdropping attack in wireless communication. 
The wiretap adversary is a passive eavesdropping adversary at the physical 
layer of communication. 

In this paper we propose a model for adversarial 
wiretap (AWTP) channel to capture active attacks at this layer. 
We consider a $(\rho_r , \rho_w)$ wiretap adversary who can see a fraction $\rho_r$, and 
modify a fraction $\rho_w$, of the sent codeword. The code components that are read 
and/or modified can be chosen adaptively, and the subsets of read and modified 
components in general, can be different. AWTP codes provide secrecy and 
reliability for communication over these channels. We give security and 
reliability definitions for these codes, and define secrecy capacity of an AWTP channel.
The paper has 
two main contributions. First, we prove a tight upper bound on the rate of 
AWTP codes 
for $(\rho_r , \rho_w)$-AWTP channels, and use the bound 
to derive the secrecy capacity of the channel.
Second, we give an explicit construction for a perfectly secure  capacity 
achieving AWTP code family. 
 We show 
that  our AWTP model is a natural generalization of Wyner's wiretap models, and 
somewhat surprisingly, also provides a generalization of a seemingly 
unrelated cryptographic primitive, Secure Message Transmission (SMT). 
This relation is used to derive a new (and the only known)  bound on the transmission rate of one round $(\epsilon, \delta)$-SMT 
protocols. We discuss our results, 
give the relations with related  primitives, and propose directions for future work.
\end{abstract}


\IEEEpeerreviewmaketitle

\section{Introduction} \label{intro}

\IEEEPARstart{W}{yner's} pioneered wiretap channel  model \cite{W75} where noise in the channel is
used as a resource 
to provide (asymptotic) perfect secrecy  against a computationally unbounded adversary,
without requiring a shared key. In Wyner's original wiretap model and its generalization \cite{CK78}, the sender is connected to the receiver by 
 a noisy channel referred to as the {\em main channel}, and the eavesdropper 
obtains a noisy view of the communication through 
a second noisy channel, referred to as the {\em wiretapper channel}. 
The goal of the model is to provide asymptotically reliable (noise free) and secure (perfect secrecy) message transmission from
Alice to Bob.
In a follow-up work Wyner (and Ozarow)
 \cite{OW84} introduced wiretap II model where the main channels is noiseless and the wiretapper channel is an erasure channel where the erasures are controlled by the adversary: the adversary  can select the subset of codeword
 components that he/she would like to see.
 The goal here is to provide perfect secrecy for the communicants.
 {\em Secrecy capacity} of a wiretap channel 
is  the fraction of a sent symbol that can be received with perfect secrecy and reliability.
Wyner derived secrecy capacity of a {\em degraded } wiretap channel  where the wiretapper channel is
a concatenation of the main channel and a second noisy channel, and showed the existence of 
codes that achieve secrecy capacity. 
Similar results for wiretap channel II,  and 
also general wiretap 
channel were obtained in  \cite{OW84} and \cite{CK78}, respectively.

Wiretap model naturally captures physical layer wireless communication where the sender's transmission can be intercepted (eavesdropped) by a third party who is within the reception distance of the transmitter.
There is a large body of research \cite{BB11,CDS12,CPFD11,LH78,LPS08,OW84,M92,MW03,MW00,MBL09,LGP08,MV11,CPFD11} on variations of the basic model including extending the goal of communication to key agreement also. There have also been  numerous implementations  of  the system \cite{BTV12,BB11}.

Considering active adversaries  in wiretap model is well motivated by real life application scenarios.
In wireless communication 
it is relatively easy for an attacker to
inject signals in the channel resulting in the transmitted symbols to be erased, or selectively modified
 \cite{PTDC11}. Studying active adversaries that model channel tampering   is also important
for bringing 
wiretap model in line  with  cryptographic models used for confidentiality (e.g. authenticated encryption). 
Recent proposals \cite{MBL09,BS13,ALCP09}
for physical layer active adversaries in wiretap setting consider a general adversary 
modelled as an arbitrary varying channel, but fall short on one or more of the following,
(i) considering adaptive adversaries that uses its current knowledge to perform its next action, (ii) using a strong definition of security, (iii) deriving an expression or a tight upper-bound for secrecy capacity,
and (iv) providing an efficient explicit construction.

In this paper we propose an {\em Adversarial Wiretap Channel  (AWTP Channel)} model 
in which the  adversary can adaptively choose his/her view of the channel, and
tamper with the transmission over the channel using this view.  Adversary's observation and tampering strategies are arbitrary  as long as the total number of observed and tampered symbols stay within specified limits. The model effectively replaces probabilistic noise over the main channel in Wyner's wiretap model with adversarial noise,
and for the wiretapper channel adopts the Wyner's wiretap II model, allowing 
 the adversary to adaptively select the symbols that 
they want to observe.
The model thus can be seen as extensions of Wyner's wiretap and wiretap II models both.
In Section \ref{sec_smtawtp} we will also show that this model  also generalizes the  seemingly
unrelated cryptographic model of Secure Message Transmission \cite{FW00,KS09,FFGV07,PCSR10}, used for networks. 
There are however subtle differences between the two that will be further discussed in  Section \ref{sec_smtawtp}.
We use a  definition of security and reliability 
that is in-line with related cryptographic primitives such as
SMT and also the security definition of wiretap adversaries in \cite{BTV12}.%

 We will derive an expression
 for upper bound on the rate of AWTP codes and 
give an explicit construction for a code that achieves the bound. 


\subsection{Our Results}

\subsubsection{AWTP Channel} An AWTP channel  is
specified  by a pair of parameters $(\rho_r, \rho_w)$:
 for a codeword of length $N$,  the adversary can choose  a subset $S_r$ of  $\rho_r N$ components of the codeword to read, and a
 subset $S_w$ of 
$ \rho_w N$ components to write, and  {\em writing is by adding an error vector to the codeword.}
The goal is to provide reliability and secrecy for communication against the above adversary.
Secrecy is defined as the indistinguishability of the adversary's view of the 
communication for  two  messages
 chosen by the
adversary, and indistinguishability is measured by the statistical distance between the two views.
Reliability is given by the receiver's probability of correctly decoding a sent message, possibly chosen by the adversary (See Definition \ref{def_awtpcode}).
{\em Perfect secrecy} and {\em  perfect reliability} correspond to  zero information leakage 
and always correct recovery\footnote{Note that similar to Hamming codes, because of the adversarial nature of error, it is possible to require perfect decoding.} of the message by the receiver, respectively. The $\epsilon$-{\em secrecy capacity} ${\mathsf C}^\epsilon$
of a $(\rho_r, \rho_w)$-AWTP is the highest possible information rate (number of message bits divided by the number of communicated bits) with $\epsilon$-secrecy and the guarantee that the decoder error probability asymptotically approaches zero.  
Secrecy capacity of a channel gives the {\em potentials} of the channel  for secure communication
 and achieving this capacity with efficient construction,
is the ultimate goal of the system designer.

\subsubsection{Adversarial Wiretap Codes (AWTP Code)} 
An AWTP code provides security and reliability for message  transmission over $(\rho_r, \rho_w)$-AWTP channels. An  AWTP code is specified by a triple $(\cM, N, \Sigma)$, denoting the message space, code length and alphabet set respectively, and a pair of algorithms  $(\mathsf{AWTPenc}(\cdot), \dec(\cdot))$ that are used for encoding and decoding, respectively. Encoding is probabilistic and maps a message $m \in \cM$ to a codeword $c\in C$.   Decoding is deterministic and outputs a message that could be incorrect. Decoding error is worst case and assumes that  the adversary uses their  best strategy for
choosing the message (message distribution) and tampering with communication to make the decoder output in error. An $(\epsilon, \delta)$-AWTP code guarantees that the information leaked about the message (measured using statistical distance) and the probability of decoding error are upper bounded by $\epsilon$ and  $\delta$, respectively.
\remove{
 {\rd An AWTP {\em  code family} $\mathbb{C^\epsilon}$ is a family of $(\epsilon,\delta)$-AWTP codes, indexed by the code length $N$. 
 The {\em rate of a code $C$} is denoted by $R(C)$ and is defined as $R(C)\stackrel{\triangle}= \frac{\log |\cM|}{ N\log |\Sigma|}$. A  {\em rate $R(\mathbb{C}^\eps)$ is achievable} by a code  family $\mathbb{C}^\epsilon $, if 
for any $\xi>0$ there exists an $N_0$ such that for all $N\geq N_0$, there is an \edwtp\; code that has rate $R(C)\geq \frac{1}{N}\log_{|\Sigma|}|\cM|-\xi$, and $\delta\leq \xi$.
}
}
 The {\em rate of an AWTP code $C^N$} of length $N$, denoted by $R(C^N)$, is defined as $R(C^N)= \frac{\log |\cM|}{ N\log |\Sigma|}$. An AWTP {\em  code family} $\mathbb{C}$ is a family $\{C^N\}_{N\in \mathbb{N}}$ of AWTP codes indexed by the code length $N$.  The {\em rate $R(\mathbb{C})$ is achievable} by a code  family $\mathbb{C}$, if for any $\xi>0$ there exists an $N_0$ such that for all $N\geq N_0$, we have $\frac{1}{N}\log_{|\Sigma|}|\cM|\geq R(\mathbb{C})-\xi$, and decoding error probability satisfies $\delta\leq \xi$. The $\epsilon$-secrecy capacity of AWTP channel, denoted by $\mathsf{C}^\epsilon$, is the largest achievable rate of all AWTP-code families that provide $\epsilon$-secrecy for the channel. ($\mathsf{C}^0$ for perfect secrecy).

\subsubsection{Rate Upper Bound of AWTP channel} 
We  prove
the bound $\bH(M)\leq \log|\Sigma|(1-\rho_r-\rho_w)N$, for an arbitrary message distribution with entropy
$\bH(M)$.
Using  uniform message distribution, we obtain  an upper bound on the  rate of  an $(\epsilon, \delta)$-AWTP code for a $\rorw$-AWTP channel, 
that 
leads to the following upper bound on the secrecy capacity of a 
$(\rho_r, \rho_w)$-AWTP channel, 
\begin{equation} \label{code-rate1}
\mathsf{C}^\epsilon\leq 1-\rho_r-\rho_w+2\epsilon \rho_r(1+\log_{|\Sigma|}\frac{1}{\epsilon})
\end{equation}
For $\epsilon=0$, we obtain the upper bound, 
$\mathsf{C}^0\leq 1-\rho_r-\rho_w,$
on the secrecy capacity of perfectly secure AWTP code families
(Corollary \ref{le_up2}), which  is achieved by the construction 
in Section \ref{sec_construction} (Theorem \ref{the_awtpcode}), and  we obtain the perfect secrecy capacity of a $\rorw$-AWTP channel,
\begin{equation}\label{code-rate}
\mathsf{C}^0= 1-\rho_r-\rho_w
\end{equation}

The bound on $\mathsf{C}^0$  implies that perfect security for $\rorw$-AWTP channels is possible 
only if, 
$\rho_r+\rho_w<1$, 
indicating a trade-off between read and write 
(adding noise) capabilities of an AWTP adversary.  In particular,  when the adversary is almost oblivious ($\rho_r$ is small), 
the rate stays positive even when the adversary writes over the large  fraction  $<1-\rho_r$ of the codeword, 
and on the other extreme when $\rho_r$ is close to 1,  few  $<1-\rho_r$ corrupted symbols can be tolerated. 
In general, the subset of components of a codeword that is either read, or written to, cannot
 contribute to secure and reliable transmission of information.  Since the capacity result 
must hold for {\em all} adversaries, that is all choices of $S_r$ and $S_w$ (subject to the bound on the size), considering 
 an adversary that uses $S_r\cap S_w = \emptyset$, results in the
capacity to be less than $1-\rho_r-\rho_w $. 
Oblivious writing adversaries have been previously considered in  Algebraic Manipulation Detection (AMD) codes \cite{CDFP08} 
where the goal is to {\em detect algebraic tampering. }
In AWTP codes however, the aim is to correct errors and recover the sent message.

\subsubsection{A Capacity Achieving AWTP Code Family} \label{sec_intro_codefamily} 
We construct a $\del$-AWTP code family $\mathbb{C}=\{C^N : N\in \mathbb{Z}\}$, where $N$ is the code length, for a $(\rho_r, \rho_w)$-AWTP channel. For any small $\xi >0$, the code $C^N$ has  rate $R(C^N)= 1-\rho_r- \rho_w-\xi$;  the code alphabet size is $|\Sigma|=\mathcal{O}(q^{1/{\xi}^2})$, and decoding error probability is $\delta \leq q^{\frac{1}{\xi}-N}$. The construction gives a construction for a code family that achieves the capacity ${\mathsf C}^0$.

The construction uses three building blocks: \emph{Algebraic Manipulation Detection Code} (AMD code), \emph{Subspace Evasive Sets}, and 
{\em Folded Reed-Solomon code (FRS code)}. An AMD code \cite{CDFP08} (Definition \ref{def_amd}) is a randomized  code that protects against an oblivious adversary that ``algebraically manipulates" (adds error to) the AMD codeword; A $(v, \ell)$-subspace evasive sets is a subset $\cal S$ of a vector space $\bF_q^n$ with the property that, any subset of dimension $v$ has at most $\ell$ common elements with $\cal S$; An FRS code \cite{Gur11} is a special type of Reed-Solomon code that has efficient list decoding algorithm for errors up to the list decoding capacity.

AWTP encoding algorithm of a message uses these three building blocks as follows:
 the message is first encoded into an AMD codeword;  the resulting AMD codeword is then encoded into 
a vector in a subspace evasive sets with appropriate parameters; the resulting vector is finally encoded into a codeword of the FRS code. AWTP decoding algorithm uses inverse steps: it  first uses the FRS decoder to output a list of possible codewords which contains the correct codeword. Using the decoding algorithm of FRS code \cite{Gur11},  the list size will be exponential in the code length $N$.
Using the intersection algorithm of the subspace evasive sets, the list is pruned to a list of   size at most $\ell$ that contains the correct codeword. The decoder combines the above two steps and so effectively avoids the generation of the 
exponential size list in the FRS decode, which would result in inefficient decoder. The final step uses the AMD code  to find the correct coded message, and outputs the correct message with probability at least $1-\delta$. We prove that with appropriate choices of parameters, the rate of the code family meets the rate upper bound of the of $(\rho_r, \rho_w)$-AWTP channels with equality, and so the family is capacity achieving.

\subsubsection{Relations with SMT and AMD codes} 
AWTP model of secure and reliable communication over adversarially controlled channels,
is closely related to 1-round SMT \cite{DDWY93}, a model proposed for secure and reliable communication in networks. In SMT setting Alice is connected to Bob through a
set of $N$ node disjoint paths  ({\em wires}) in a network, $t$ of which are controlled by a Byzantine adversary. The goal of an SMT protocol is to provide reliability and privacy for communication: an $(\epsilon, \delta)$-SMT ensures that the privacy loss (indistinguishability based) is bounded by $\epsilon$, and 
the probability of failing to decode the message is bounded by $\delta$ \cite{KS09}.
A notable difference between an AWTP channel adversary and an SMT adversary is that, 
in the former error is {\em added to the codeword} while in the latter, it is {\em a replacement error} and 
allows the adversary to replace what is sent over a wire with  its own adversarial choices. In Section \ref{sec_smtawtp}, we consider the relationship between the two primitives, and in particular show how the results in  one, can give results for the other. The relationship also suggests 
a new efficiency measure for SMT systems using information rate of a family.


\subsection{Related Work}\label{sec_relatedwork}

Wiretap model and its extensions has attracted much attention in recent years.
There is a  large body of excellent  works on extensions of wiretap model  \cite{CK78, OW84, M92, MW03,LGP08, CDS12},  construction of capacity
achieving codes  \cite{HM10, BTV12}, and implementation of codes in practice \cite{BBRM08,BB11} . We only consider the works that are directly related to this work. Considering adversarial control in wiretap channel dates back to Wyner wiretap II \cite{OW84} model in which the adversary can select their view of the communication. The adversary  however, does not modify the transmission over the main channel which is assumed noise-free. 
Physical layer  active adversaries for wiretap channels that tamper with the transmission,
 have been considered  more recently
 \cite{MBL09,BS13,ALCP09}. 
These works model  wiretap channels
under active attack, as an  arbitrarily varying channel. 
An {\em arbitrarily varying  channel  (AVC)} \cite{BBT60,A78,CN88, BBS13} is 
  specified by two finite sets $\cX$ and $\cY$ of input and output alphabets, a finite set $\cA$ of
 {\em channel states},
 and  a set of channels specified by transition probabilities $\bP(y|x, a), x\in \cX,  y\in \cY, a\in\cA$. The channel state in general varies with each channel use (possibly with memory) and, \[ \bP(y^n|x^n, a^n) = \Pi_{i=1}^n \bP(y_i| x_i,a_i)\]
where $a^n= (a_1\cdots a_n)$, $a^n\in {\cal A}^n$, is the sequence of channel states.
An {\em 
 arbitrarily varying wiretap channels (AVWC)} is specified by an input alphabet set $\cX$,  two sets of output
 alphabets, $\cY$ and $\cZ$, representing the {\em legitimate receiver's} and {\em the wiretapper's} input values, respectively,
 and a family of channels, each specified  with a transition probability $\bP(y,z|x, a), x\in \cX, y\in \cY, z\in \cZ, a\in \cA)$
  indexed by the channel state $a$.
In  \cite{MBL09},  
t
a {\em jammer}  chooses the state $a_i$ (jamming signal)
 independent of the eavesdroppers' observation $z$. 
Transmitter and receiver know the state space but not the state chosen
by the adversary.\\
The message is chosen randomly, with uniform distribution, from the message space.
{\em Encoding and decoding is  randomized}; that is the system uses a family of encoder and decoder  pairs,  and the pair used by the sender and receiver is specified by a random 
value (also called key) that is known to the eavesdropper but {\em not the jammer}. The family of
the codes is known to the jammer.
Security is measured by the {\em rate of the mutual information between the message and the adversary's observation}, and
reliability
for  a single encoder and decoder is in terms of the 
 expected error probability over all messages. For randomized codes
security and reliability  are averaged over all realizations of the code.\\
Authors defined randomized-code  secrecy capacity of AVWP channels and 
derived an upper bound for that. They also obtained 
the capacity
for the special case of {\em strongly
degraded with independent states} where  certain Markov chains hold among  $X, Y$ and $Z$,
and the set of states $\cA$ is decomposable as ${\cal A} = {\cal A}_y \times {\cal A}_z $ 
and states of the receiver's channel and the eavesdroppers channels are selected 
independently.\\
In summary,  the model (i) assumes common randomness between the sender and the receiver that is unknown to the jammer,  but is known to the eavesdropper,
(ii) uses weak definition of secrecy,  and (iii)  the jammer's corruption does not depend on the eavesdropper view. 
 In \cite{BBS13, BS13} these results are strengthened. These works both use a  strong definition of secrecy 
using total mutual information (instead of rate),  and  \cite{BS13} uses the AVWC model and analyzes active adversaries that exploit common randomness.

Our adversarial channel model can be seen as a special class of arbitrarily varying channel.  For this class  we can remove some of the restrictions of
the above line of work.  More specifically, 
 (i) we do not assume  shared randomness between the sender and
the receiver, (ii)  we consider an integrated adaptive eavesdropping and jamming adversary and
assume 
that  to corrupt the next symbol, the 
adversary uses all its knowledge up to that point; (iii)  we allow adversary to choose the  message distribution
and so the error is worst case;
and finally (iv)
 the secrecy measure in our case is in terms of statistical distance between the adversary's views of two adversarially chosen
 messages. We note that in \cite{BS13}, secrecy is measured as
 the mutual information of a {\em random message}  (uniform distribution on messages) and the adversary's view.   
Our security definition using statistical distance  is equivalent  to {\em the mutual information security when the message distribution is adversarially chosen} \cite{BTV12}.

In \cite{ALCP09}  wiretap II model  is extended to include an active adversary, using two types of corruption.
In the first one the adversary 
erases symbols that are observed, and in the second, corrupts them. 
Authors give constructions that 
 achieve good rates.
The adversary's corruption  capability in our work is more general (adversary is more powerful). 
No other comparison can be made because of insufficient details.

Wyner \cite{W75} quantifies security of the system using the adversary's  equivocation defined as the average (per message symbol) uncertainty about the message, 
given the adversary's view of the sent  codeword.
Strengthening this security definition has been considered in \cite{MW00,CDS12,HM10,BTV12}. In  \cite{BTV12}, 
 the relationship  among security notions used for wiretap channels  is studied, and it is also shown that   distinguishability based definition using statistical distance is equivalent to a 
  security notion  that is called {\em mutual information security}. This  is  a stronger notion compared to
 the strong security notion used in
  \cite{MW00}, with the difference being that the 
 adversary chooses the message distribution.

Adversarial channels have been widely studied in literature \cite{CN88,GS10,M08},
with \cite{LN98} providing a comprehensive survey. 
An adversarial channel closely related to this work  is {\em limited view (LV) adversary channel } \cite{SW13,SW131}.  
An LV adversary power is identical to 
the adversary in AWTP channel but the goal of communication  in the former is reliability, while transmission over AWTP channels requires 
reliability and secrecy both.

Computationally limited active adversariy at network layer of communication,  
 has been considered in \cite{M92,MW03,DKKRS12,KR09,RW04}.  This adversary can  tamper with the whole message, and to provide protection, 
 access to resources such as shared randomness
\cite{GMS74, S84},  close secrets \cite{KR09}, or extra channels \cite{NSS06}, is required.
 The adversary in AWTP setting is at physical layer of communication, 
 and the only advantage of 
 communicants over the adversary is limited access of the adversary to the channel. 
 
In all above models, the reliability goal is to correctly recover  the sent message. 
A less demanding reliability goal in  {\em detection} of errors.
Adversarial tampering by an adversary that cannot ``see" the encoded message, has been considered in the context of AMD codes \cite{CDFP08}. 
AMD codes 
with strong security, are  randomized codes that detect algebraic manipulation resulting from the addition of an arbitrary error vector.
Weak AMD codes are deterministic and use the message randomness 
to provide protection. 
AMD codes with leakage \cite{AS13} allow the adversary to ``see" a fraction of the codeword. The writing ability of the adversary however is unrestricted. This is possible because the goal  of the encoding is to detect tampering, while in  AWTP model, the goal is message recovery and so the corruption must be limited.


\subsection{Discussion and Future Work}
Providing security against a computationally  unlimited adversary and  without 
assuming a shared key,  requires limiting adversary's physical access to the information.
\remove{
In AWTP model, the adversary can choose their view of the codeword, and apply arbitrary (additive) modifications as long
as the fractions of read and modified components, are bounded. 
}
The bound on the write ability of the adversary captures  limitations of the  adversary's transmitting power,
 and  the complexity of effecting symbol change \cite{PTDC11} in real systems.
The bound on the read ability captures limited physical access to sent symbols  due to 
the inadequacy of the adversary's receiver to perfectly decode all the sent symbols. 
\remove{Our formulation of the adversarial read and write capabilities, allows us to model a range of adversaries 
with different capabilities.

Compared to previous models of adversarial wiretap channel 
{\rd using arbitrary varying channels,  although our model is 
less general, but it considers an adaptive adversary while in the former, 
separate eavesdropping and jamming adversaries with no communication between them is considered.}

Although in our general formulation of the problem  the adversary's jamming signal
 is {\em added to the communicated codeword},  by limiting the adversary's corrupted symbols to
be a subset of the observed symbols, this restriction can be removed. That the adversary
can apply any corruption to the read symbols.
These restricted-adversaries give a one-to-one correspondence between AWTP codes and
SMT in the context of networks. \\
}

A number of results in this paper can be improved.  
Construction of capacity achieving AWTP codes for $\epsilon> 0$, 
and construction of  AWTP code for constant size alphabets, and in particular 
$\mathbb{F}_2$,  remain open problems.

\vspace{1.5mm}
\noindent
{\em Organization:}
Section 2 provides background. 
In Section 3, we introduce AWTP channels and  codes. 
Section 4 is on bounds and Section 5 gives the construction.
Section 6 concludes the paper.


\section{Preliminaries}


We use  calligraphic symbols $\cal X$ to denote sets, $\bP(X) $ to  denote a probability distribution over  ${\cal X}$, and $X$ to denote a random variable that takes values from  $\cal X$ with probability $\bP(X)$.  
The conditional probability given an events $E$ is 
$\bP[X=x | E]$. 
$\log(\cdot)$ is  logarithm in base two.
{\em Shannon entropy } of a random variable $X$ is, 
$\bH(X)=-\sum_{x}\bP(x)\log \bP(x)$, and 
{\em  conditional entropy} of a random variable $X$ given $Y$ is defined by,
$\bH(X | Y)=-\sum_{x,y}\bP(x, y)\log \bP(x|y).$
{\em Statistical distance} between two random variables $X_1$, $X_2$ defined over the same set is given by
$\SD(X_1, X_2)=\frac{1}{2}\sum_{x}|\bP[X_1=x]-\bP[X_2=x]|$.
{\em Mutual information} between random variables $X$ and $Y$ is given by,
$
\bI(X, Y)= \bH(X)-\bH(X|Y)
$.
For a vector $e$, {\em Hamming weight} of a vector $e$ is denoted by be $wt(e)$.

\subsubsection{Algebraic Manipulation Detection Code (AMD code)}\label{sec_amd}
Consider a storage device $\Sigma({\cal G})$ that  holds an element $x$ from a group $\cal G$. The storage $\Sigma({\cal G})$ is private
 but can be manipulated by the adversary by adding  $\Delta\in {\cal G}$.  AMD codes allow the manipulation to be detected.

\begin{definition}[AMD Code\cite{CDFP08}]\label{def_amd}
An $({\cal X}, {\cal G}, \delta)$-Algebraic Manipulation Detection code $($$({\cal X}, {\cal G}, \delta)$-AMD code$)$ consists of two algorithms $(\mathsf{AMDenc}, \mathsf{AMDdec})$. 
Encoding, $\mathsf{AMDenc}: {\cal X} \rightarrow {\cal G}$,  is probabilistic and   maps an element of a set $\cal X$ to an element of an additive group $\cal G$. 
Decoding,  $\mathsf{AMDdec}: \cal G \rightarrow {\cal X} \cup \{\perp\}$, is deterministic and for any $x \in {\cal X}$, we have $\mathsf{AMDdec}(\mathsf{AMDenc}(x)) = x$.  Security of AMD codes is defined by requiring, 
\begin{eqnarray} \label{amd}
\bP[\mathsf{AMDdec}(\mathsf{AMDenc}(x) + \Delta) \in \{x,\perp \}] \leq \delta, \,\, 
\end{eqnarray}
for all $x\in {\cal X}, \Delta \in {\cal G}.$
\end{definition}

An  AMD code is {\em systematic} 
if  the encoding has the form $\mathsf{AMDenc} : {\cal X}\rightarrow {\cal X} \times {\cal G}_1 \times {\cal G}_2$, and $x\rightarrow (x, r, t=f (x, r))$,
for some function $f$ and $r \stackrel{\$}\leftarrow {\cal G}_1$. 
The decoding function results in $\mathsf{AMDdec}(x, r, t) = x$, if and only if $t=f (x, r)$, and $\perp$ otherwise.

We use the systematic AMD code in \cite{CDFP08} over an extension field. 
Let $\phi$ be a bijection between vectors $\bf v$ of length $N$  over $\bF_q$,
 and elements of $\bF_{q^N}$, and let $d$ be an integer such that $d + 2$ is not divisible by $q$. Define the function $\mathsf{AMDenc}: \bF_{q^N}^d \rightarrow \bF_{q^N}^d \times \bF_{q^N} \times \bF_{q^N}$ as, $\mathsf{AMDenc}(x) = (x,r,f(x,r))$ where,
\[
f(x, r)=\phi^{-1}\left(\phi(r)^{d+2}+\sum_{i=1}^{d}\phi(x_i)\phi(r)^i\right)\mod q^N
\]

\begin{lemma}\label{le_amd}
For the AMD code above, the success chance of an adversary, 
that  has no access to the  codeword $(x,r,t)$, in constructing a new codeword  $(x',r',t')= (x'=x+\Delta x, r'=r+\Delta r, t' =t+\Delta t)$,
that  satisfies $t' = f(x', r')$, is no more than $\frac{d+1}{q^N}$. 
\end{lemma}

Proof of this Lemma is the direct application of Theorem 2 in \cite{CDFP08}, when the underlying field is $\bF_{q^N}$.

\subsubsection{Subspace Evasive Sets}\label{sec_ses}


Subspace evasive  sets are used to reduce the list size of list decodable code \cite{DL12}.

\begin{definition}[Subspace Evasive Sets\cite{DL12,Gur11}]
Let $\mathcal{S}\subset \bF_q^n$. We say $\mathcal{S}$ is a $(v, \ell)$-subspace evasive if for all $v$-dimensional affine subspaces $\mathcal{H}\subset \bF_q^n$, we have $|\mathcal{S}\cap \mathcal{H}|\leq \ell$.
\end{definition}

Dvir {\em et al.} \cite{DL12} gave an efficient explicit construction of subspace evasive sets $\cS\subset \bF_q^{n}$, with an efficient {\em intersection algorithm} that computes $\cS \cap \cH$ for  any $v$-dimensional subspace $\cH\subset \bF_q^n$. 

\remove{
Let $\bF$ be a field and $\overline{\bF}$ be its algebraic closure. A variety in $\overline{\bF}^w$ is the set of common zeros of one or more polynomials. Given $v$ polynomials $f_1, \cdots, f_v\in \overline{\bF}[x_1, \cdots, x_w]$, we denote the variety as
\[
{\bf V}(f_1,\cdots, f_v)=\{{\bf x}\in \overline{\bF}^w\;|\;f_1({\bf x})=\cdots=f_v({\bf x})=0\}
\]
where ${\bf x}=\{x_1, \cdots, x_w\}$.

For a polynomials $f_1, \cdots, f_v\in \bF[x_1, \cdots, x_{w}]$, we define the common solutions in $\bF^w$ as
\[
\begin{split}
{\bf V}_{\bF}(f_1,\cdots, f_v)&={\bf V}(f_1,\cdots, f_v)\cap \bF^w\\
&=\{{\bf x}\in \bF^w\;|\;f_1({\bf x})=\cdots=f_v({\bf x})=0\}
\end{split}
\] 
}

A $v\times w$ matrix is called {\em strongly-regular}  if all its $r\times r$ minors are regular (have non-zero determinant) for all $1\leq r\leq v$.  
\begin{lemma}(Theorem 3.2 \cite{DL12})
Let $v\geq 1, \varepsilon>0$ and $\bF$ be a finite field. Let $w=v/\varepsilon$ and,  assume $w$ divides $n$. Let $A$ be a $v\times w$ matrix with coefficients in $\bF$ which is strongly-regular. Let $d_1>\cdots>d_w$ be integers. For $i\in [v]$ let,
\[
f_i(x_1,\cdots, x_w)=\sum_{j=1}^wA_{i,j}x_j^{d_j}
\]
and define the subspace evasive set $\cS\in \bF^n$ to be $(n/w)$ times cartesian product of ${\bf V}_{\bF}(f_1, \cdots, f_v)\subset \bF^w$. That is,
\[
\begin{split}
\cS&={\bf V}_{\bF}(f_1, \cdots, f_v)\times \cdots \times {\bf V}_{\bF}(f_1, \cdots, f_v)\\
&=\{{\bf x}\in \bF^n: f_i(x_{tw+1},\cdots, x_{tw+w})=0,\\
&\qquad\qquad\qquad \forall 0\leq t<n/w, 1\leq i\leq v\}
\end{split}
\]
Then $\cS$ is $(v, v^{D\cdot v\log\log v})$-subspace evasive set, and $|\cS|=|{\bF}|^{(1-\varepsilon)n}$.
\end{lemma} 

To use a subspace evasive sets for efficient list decoding, two efficient  algorithms are needed:
(i) a bijection mapping that   maps messages that are 
elements of a space $\bF^{n_1}$  into the subspace evasive set $\cS$,  and an intersection algorithm that computes the 
intersection between $\cS$ and any subspace $\cH$ with dimension at most $v$. The lemmas below show the existence of these two algorithms for the subspace evasive set above. 

\begin{lemma}\cite{DL12}\label{le_sesencoding}
Let $v, w, n_1\in \mathbb{N}$, $b=\frac{n_1}{w-v}$, $n=bw$, and $\bF_q$ be a finite field. For any vector ${\bf v}\in \bF_q^{n_1}$, there is a bijection which maps $\bf v$ into an element  of the subspace evasive set ${\cal S}\subset \bF_q^n$. That is,  $\mathsf{SE}: {\bf v}\rightarrow {\bf s}\in \cS$. The encoding algorithm is $\pl(n)$.
\end{lemma}


\begin{lemma}\cite{DL12}\label{le_sesdecoding}
Let $\cS\subset \bF_q^n$ be the $(v, \ell)$-subspace evasive sets (described above). Then there exists an algorithm that, given a basis for any $\mathcal{H}$, outputs $\cS\cap \mathcal{H}$ in time $\pl(v^{v\cdot \log\log v})$.
\end{lemma}


\subsubsection{Folded Reed-Solomon Code (FRS code)} \label{FRS_Code)}

An error correcting code $C$ over $\bF_q$ is a subspace of $\bF_q^N$.
The rate of the code is $\log_2 |C|/N$.
A code $C$ of length $N$ and {\em rate} $R$ is $(\rho, \ell_{\mathsf{List}})$-list decodable if the number of codewords within distance $\rho N$  from any received word is at most $\ell_{\mathsf{List}}$. 
List decodable codes can  correct up to 
$1-R$ fraction of errors in a codeword, which is twice 
that of unique decoding. This is however at the cost of outputting a list of possible sent codewords (messages).
Construction of  good codes with efficient list decoding algorithms has been an active research area.
An explicit construction of a list decodable code that achieves the list decoding capacity $\rho=1-R$,  
is given by Guruswami et al. \cite{Gur11}. The code is called {\em Folded Reed-Solomon code (FRS code)}, and can be seen as a Reed-Solomon code with extra structure. 
The code has polynomial time encoding and list decoding algorithms.

\begin{definition}\cite{Gur11}
A $u$-Folded Reed-Solomon code is an error correcting  code with block length $N$ over $\bF_q^u$ where $q>Nu$. The message of an FRS code is written
as a polynomial $f(x)$ with degree $k$ over $\bF_q$. 
The FRS codeword corresponding to the message is a vector over $\bF^u_q$ where each component is a $u$-tuple $(f(\gamma^{ju}), f(\gamma^{ju+1}),\cdots, f(\gamma^{ju+u-1}))$, $0 \leq j < N$, and $\gamma$ is a generator of $\bF_q^*$, the multiplicative group of $\bF_q$. A codeword of a $u$-folded Reed-Solomon code of length $N$ is in one-to-one correspondence with a codeword $c$ of a Reed-Solomon code of length $uN$, and is obtained by grouping together $u$ consecutive components of $c$.
We use $\mathsf{FRSenc}$ to  denote the encoding algorithm of the FRS code. 
$u$ is called the {\em  folding parameter} of the FRS code. 
\end{definition}

We will use the {\em linear algebraic FRS decoding algorithm}  of these codes \cite{Gur11}\remove{that is outlined in the appendix \ref{decode_FRS}}\remove{(Appendix \ref{decode_FRS})}. 

\begin{lemma} \cite{Gur11} \label{le_fd} For a Folded Reed-Solomon code of block length $N $ and rate $R = \frac{k}{uN}$, the following holds for all integers $1\leq v\leq u$. Given a received word $y \in (\bF_q^{u} )^N$ agreeing with $c$ in at least a fraction,
\[
N-\rho N>N(\frac{1}{v+1}+\frac{v}{v+1}\frac{uR}{u-v+1})
\]
one can compute a matrix ${\bf M} \in \bF_q^{k\times (v-1)}$ and  a vector ${\bf z} \in \bF^k_q$,  such that the message polynomials $f \in \bF_q[X]$ in the decoded list are contained in the affine space ${\bf M}{\bf b}+{\bf z}$ for ${\bf b} \in \bF^{v-1}_q$ in $O((Nu\log q)^2)$ time.
\end{lemma}


\section{Model and Definitions}\label{sec_awtp}
We consider the following scenario. 
Alice wants to a send messages $m \in {\cal M}$, reliably and securely to Bob, over  a communication channel that is partially controlled by an adversary, Eve. Let $\Sigma$ denote the channel  alphabet, and $\cC$ be a code,  $\cC \subset \Sigma^N$, together with a {\em probabilistic  encoder, }  $\enc : {\cal M}\times {\cal R} \rightarrow \cC,$  and a  {\em deterministic decoder},  $\dec :  \Sigma^N  \rightarrow {\cal M}$. The  encoder takes a message $m$ and a random string   $r_{\cal S} \stackrel{\$}\leftarrow \cal R$ and outputs a codeword $c= \enc(m, r_{\cal S})$. The codewords associated with a message $m$  and  different $r_{\cal S}$, define a random variable  over  $\cC$. Alice will use the encoding algorithm and for a message $m$ (also referred to as the {\em information }), generates 
a codeword $\enc(m)$. The adversary interacts with the codeword as described below, resulting  Bob to receive a corrupted word $y\neq c$. Bob uses the decoding algorithm to recover the message.

\subsection{Adversarial Wiretap:  Channel and Code}
 Let $[N]=\{1,\cdots, N\}$, and  $S_r= \{i_1,\cdots, i_{\rho_rN}\} \subseteq [N]$ and $S_w= \{j_1,\cdots, j_{\rho_wN}\} \subseteq [N]$, denote two subsets of the $N$ coordinates, and for a vector $x\in \Sigma^N$, $\mathsf{SUPP}(x)$   denote the set of coordinates where  $x_i$ is non-zero.

\begin{definition}\label{def_awtpchannel}
A {\em$(\rho_r, \rho_w)$-Adversarial Wiretap channel $($or a $(\rho_r, \rho_w)$-AWTP channel$)$}, is an adversarially corrupted
 communication channel between Alice and Bob, such that it is (partially) controlled by the adversary Eve with two capabilities: Reading and Writing. For a codeword of length $N$, Eve can do the following.
\begin{itemize}
\item {\bf Reading} (also called {\em Eavesdropping}):  Eve can select a subset $S_r\subseteq [N]$ of size at most $\rho_rN$ and read the components of the sent codeword $c$, on positions associated with $S_r$. Eve's view of the codeword is given by, 
$
\mathsf{View}_{\cal A}(\enc(m, r_{\cal S}), r_{\cal A})=\{c_{i_1},\cdots, c_{i_{\rho_rN}}\},
$
and consists of all the components that are read (observed).
\item {\bf Writing} (also called {\em Jamming}): Eve  can choose a subset $S_w\subseteq [N]$ of size  at most $\rho_wN$, for ``writing". This is by adding an error vector $e$ to vector  $c$, where  the addition is component-wise over $\Sigma$. It holds that $\mathsf{SUPP}(e) \subseteq S_w$.  The corrupted components of $c$ are  $\{y_{j_1},\cdots, y_{j_{\rho_wN}}\}$  and $y_{j_\ell}=c_{j_\ell}+e_{j_\ell}$. The error $e$ is generated according to the Eve's 
best strategy for making  Bob's decoder to output in error.
\end{itemize}
\end{definition}
We assume the adversary is {\em adaptive} and can select  components of the sent codeword for reading and writing one by one, at each step using their knowledge of the codeword at that time. 

Let $S=S_r\cup S_w$ denote the set of codeword components  that the adversary either reads,  or writes to. 
We have $|S|=\rho N$,  and $\rho \leq  \rho_r+\rho_w$. 

An AWTP channel is  called \emph{ restricted} if, 
$S_r=S_w$.  Restricted AWTP channel are a special type of AWTP channel where the adversary is limited in its
selection of $S_r$ and $S_w$, and so is a weaker type of AWTP channel.

Alice and Bob will use an {\em Adversarial Wiretap Code} to provide  secure and reliable communication over AWTP channel.

\begin{definition}\label{def_awtpcode}
An  {\em $(\epsilon, \delta)$-Adversarial Wiretap Code} ($(\epsilon, \delta)$-AWTP code) over $(\rho_r, \rho_w)$-AWTP channel, consists of a randomized encoding $\enc : {\cal M}\times {\cal R} \rightarrow \cC$, from the message space $\cM$ to a code $\cC$, and a deterministic decoding algorithm $\dec:  \Sigma^N \rightarrow {\cal M}$. 
The code guarantees the following two properties:

\begin{itemize}
\item {\bf Secrecy:}  For any two messages $m_1, m_2 \in \cM$, the statistical distance between the  adversary's views, when the same randomness $r_{\cal A}$  is used by the adversary, is bounded by $\epsilon$. That is, 
\[
\begin{split}
&\mathsf{Adv}^{\mathsf{ds}}(\enc, \mathsf{View}_{\cal A})\stackrel{\triangle}=\\
&\qquad\max_{m_0, m_1}\SD(\mathsf{View}_{\cal A}
(\enc(m_1), r_{\cal A}),\\
&\qquad\qquad\qquad\;\;  \mathsf{View}_{\cal A}(\enc(m_2), r_{\cal A})) \leq \epsilon
\end{split}
\]

\item {\bf Reliability:} For any message $m$ that is encoded to $c$ by the sender, and corrupted to $y= c+e$ by the $(\rho_r,\rho_w)$-AWTP channel, the probability that the receiver outputs the correct information $m$ is at least $1-\delta$. That is,
\[
\bP(M_\cS\neq M_\cR)\leq \delta
\]
where the probability is over the choice of the message, randomness of the communicants and the adversary. 
\end{itemize}
\end{definition}

The AWTP code is {\em perfectly secure} if $\epsilon=0$.

For  $\epsilon>0$, an {\em $\epsilon$-secure AWTP code family} $\mathbb{C}^\eps$, is a  family $\{C^N\}_{N\in \mathbb{N}}$ of $(\epsilon, \delta)$-AWTP codes,  indexed by $N \in \mathbb{N}$, for a $(\rho_r, \rho_w)$-AWTP channel. When $\epsilon=0$, the  family is called  a  {\em perfectly secure AWTP code family.} 

{\em In the following, when $\eps\neq 0$, we omit $\eps$ and simply write $\mathbb{C}$ to denote, $\mathbb{C}^\eps$.}

\begin{definition}\label{def_awtpfamily}
For a  family $\mathbb{C}$ of \edwtp\;codes the {\em rate $R(\mathbb{C})$ is achievable}
if for any $\xi$, there exists  $N_0$ such that for any $N\geq N_0$, we have, $\frac{1}{N}\log_{|\Sigma|}|{\cal M}|\geq R(\mathbb{C})-\xi$, and the decoding error probability satisfies, 
$\bP(M_\cS\neq M_\cR)\leq \xi.$
\end{definition}

 To define secrecy capacity of an AWTP channel, we will use achievable rate of  a code family for the channel.
\begin{definition}\label{def_secrecycapacity}
The {\em $\epsilon$-secrecy  capacity of a $(\rho_r, \rho_w)$-AWTP channel denoted by ${\mathsf C}^\epsilon$}, is the largest achievable rate of all AWTP-code families $\mathbb{C}$ that provide  $\epsilon$-secrecy for the channel. 

The {\em perfect secrecy  capacity  of a $(\rho_r, \rho_w)$-AWTP channel} is denoted by ${\mathsf C}^0$,  and is  the highest achievable rate of perfectly secure  AWTP-code families 
 for the channel. 
\end{definition}


\section{A Bound on the Rate of \edwtp~Codes }\label{sec_upbound}
We derive an upper bound on the rate of AWTP codes, and use it to find the secrecy capacity of AWTP channels.

The bound is derived on the rate of an arbitrary code when the adversary uses a special strategy, given below.
Since the strategy can always be used, it follows that the code rate cannot be higher than the bound. 
The adversary's strategy is a probabilistic strategy.

\begin{enumerate}
\item Before the start of the transmission, the adversary selects two pairs of read and write sets, $\{S_r^i, S_w^i\}, i=1,2$, satisfying $S_r^1\cap S_w^2=\emptyset$. 
  
The adversary then selects one of the two pairs  with probability 1/2;  that is,
 $\bP(S_r^1, S_w^1)=\bP(S_r^2, S_w^2)=\frac{1}{2}$.
The set sizes satisfy the following: for $i=1,2$, we have  $|S_r^i|=\rho_rN$, $|S_w^i|=\rho_wN$, and $|S_r^i\cup S_w^i|=\rho N$,  where $0\leq \rho\leq 1$.

\item For the chosen read and write pair, $\{S_r^i, S_w^i\}$, the adversary, (i)  reads the 
 $\rho_rN$ components of the codeword corresponding to the subset $S_r^i$, (ii)  
chooses an error vector $E_i\in \Sigma^{\rho_wN}$ randomly with uniform distribution, and adds it component-wise to the codeword components corresponding to $S_w^i$.

\item The adversary chooses the uniform distribution on the message space.
\end{enumerate}

We associate a random variable $C_i$ to  the $i^{th}$ component of an $(\epsilon, \delta)$-AWTP code.  For $i=1,2$, let $C_{S_r^i}$ and $C_{S_w^i}$ be the components of a codeword on the sets $S_r^i$ 
$S_w^i$, respectively.

Let $Y$ denote the word that Bob receives.  

In the following we will derive the secrecy capacity, ${\mathsf C}^\epsilon$, of a $(\rho_r, \rho_w)$ AWTP channel.

\begin{theorem}\label{the_upper1}
 The upper bound on the  secrecy capacity of AWTP code family over $(\rho_r, \rho_w)$-AWTP channel is,
\[
{\mathsf C}^\epsilon \leq 1-\rho_r-\rho_w+2\epsilon \rho_r (1+\log_{|\Sigma|}\frac{1}{\epsilon})
\]
\end{theorem}

We first prove an upper-bound on the rate of an \edwtp\; code (Lemma (\ref{le_upper1})), and then extend the bound to the achievable rate of a code family, and so
the secrecy capacity of the channel.
To prove the bound on the rate of a code,  we prove two lemmas that use 
the secrecy and reliability guarantees of the code, respectively, and use them to prove the bound on the rate of the code.

\begin{lemma}\label{le_up4}
An $(\epsilon, \delta)$-AWTP code for a $(\rho_r,\rho_w)$-AWTP channel satisfies,
\[
\bH(M)-\bH(M|C_{S^1_r})\leq 2\epsilon\rho_rN\log\frac{|\Sigma|}{\epsilon}
\]
\end{lemma}

\begin{IEEEproof}
From the definition of $\epsilon$-secrecy we have,
\begin{equation}
\begin{split}
&\mathsf{Adv}^{\mathsf{ds}}(\enc, \mathsf{View}_{\cal A})\\
&=\frac{1}{2}\sum_{c_{S^1_r}}|\bP(c_{S^1_r}|m_0)-\bP(c_{S^1_r}|m_1)|\\
& +\frac{1}{2}\sum_{c_{S^2_r}}|\bP(c_{S^2_r}|m_0)-\bP(c_{S^2_r}|m_1)|\leq \epsilon
\end{split}
\end{equation}
This implies that for any pair of messages, $m_0, m_1\in \cM$, we have,
\[
\frac{1}{2}\sum_{c_{S^1_r}}|\bP(c_{S^1_r}|m_0)-\bP(c_{S^1_r}|m_1)|\leq \epsilon
\]
and so it follows that, 
\[
\begin{split}
&\SD(P_{C_{S^1_r}M}, P_{C_{S^1_r}}P_M)\\
&=\frac{1}{2}\sum_{c_{S^1_r}}|\bP(c_{S^1_r}|m)-\bP(c_{S^1_r})|\\
&=\frac{1}{2}\sum_{c_{S^1_r}}|\bP(c_{S^1_r}|m)-\sum_{m'\in \cM}\bP(c_{S^1_r}|m')\bP(m')|\\
&=\frac{1}{2} \sum_{c_{S^1_r}}  |\bP(c_{S^1_r}|m)-\sum_{m'\in \cM}\bP(c_{S^1_r}|m')\bP(m')|\\
&\leq\sum_{m\in \cM}\bP(m)\frac{1}{2}\sum_{c_{S^1_r}}|\bP(c_{S^1_r}|m)-\sum_{m'\in \cM}\bP(c_{S^1_r}|m')|\\
&\leq \sum_{m\in \cM}\bP(m)\epsilon\\
&=\epsilon
\end{split}
\]
By Theorem 17.3.3 (Page 370, \cite{CT06}), for  sufficiently small $\epsilon$
\[
\begin{split}
&I(M, C_{S_r^1})\\
&\leq 2\SD(P_{C_{S^1_r}M}, P_{C_{S^1_r}}P_M)\log \frac{|\Sigma|^{\rho_rN}}{\SD(P_{C_{S^1_r}M}, P_{C_{S^1_r}}P_M)}\\
&\leq 2\epsilon\rho_rN\log\frac{|\Sigma|}{\epsilon}
\end{split}
\]
\end{IEEEproof}

\begin{lemma}\label{le_up5}
An $(\epsilon, \delta)$-AWTP code for a $(\rho_r, \rho_w)$-AWTP channel, satisfies,
\[
\bH(M|YS_r^2S_w^2)\leq \bH(\delta)+\delta N\log |\Sigma|
\]
\end{lemma}

\begin{IEEEproof}
From Fano's inequality (Theorem 2.10.1, Page 38, \cite{CT06}), the decoding error probability $\delta$, implies,
\[
\begin{split}
\bH(M|Y) \leq \bH(\delta)+\delta N\log |\Sigma|
\end{split}
\]
We have  $\bP(M)= \bP( M S_r^1 S_w^1) +\bP( M S_r^2 S_w^2)$ and so,
\[
\begin{split}
&\bH(M|Y)=\bH(MS_r^1S_w^1|Y)+\bH(MS_r^2S_w^2|Y)\\
&=\bH(S_r^1S_w^1|Y)\bH(M|YS_r^1S_w^1)+\bH(S_r^2S_w^2|Y)\bH(M|YS_r^2S_w^2)\\
&\overset{(1)}{=}\bH(S_r^1S_w^1)\bH(M|YS_r^1S_w^1)+\bH(S_r^2S_w^2)\bH(M|YS_r^2S_w^2)\\
&\overset{(2)}{=}\frac{1}{2}\bH(M|YS_r^1S_w^1)+\frac{1}{2}\bH(M|YS_r^2S_w^2)\\
\end{split}
\]
(1) is because $\{S_r^i, S_w^i\}$ are selected before transmission starts, and independent of $Y$; (2) is from $\bH(S_r^1S_w^1)=\frac{1}{2}$. Since $\bH(M|YS_r^1S_w^1)\geq 0$, we have,
\[
\bH(M|YS_r^2S_w^2)\leq 2 \bH(\delta)+2\delta N\log |\Sigma|
\] 
\end{IEEEproof}

We denote the $(\epsilon, \delta)$-AWTP code with length $N$ as $C^N$, and the rate of $(\epsilon, \delta)$-AWTP code as $R(C^N)$.

\begin{lemma}\label{le_upper1}
The upper bound of rate of $(\epsilon, \delta)$ AWTP code $C^N$ over $(\rho_r, \rho_w)$ AWTP channel is,
\[
R(C^N)\leq 1-\rho_r-\rho_w+4\bH(\delta)+2\epsilon \rho_r(1+\log_{|\Sigma|}\frac{1}{\epsilon})
\]
\end{lemma}

\begin{IEEEproof}
We have,
\begin{equation}
\begin{split}
\bH(M)&=\bI(M; YS_r^2S_w^2)+\bH(M|YS_r^2S_w^2)\\
&\qquad-\bI(M;C_{S_r^1})+\bI(M;C_{S_r^1})\\
& \leq \bI(M; YC_{S_r^1}S_r^2S_w^2)-\bI(M;C_{S_r^1})\\
&\qquad+\bH(M|YS_r^2S_w^2)+\bI(M;C_{S_r^1})\\
&\overset{(1)}{=} \bI(M; YC_{S_r^1}S_r^2S_w^2)-\bI(M;C_{S_r^1}S_r^2S_w^2)\\
&\qquad+\bH(M|YS_r^2S_w^2)+\bI(M;C_{S_r^1})\\
&= \bI(M; Y|C_{S_r^1}S_r^2S_w^2)+\bH(M|YS_r^2S_w^2)+\bI(M;C_{S_r^1})\\
&= \bH(Y|C_{S_r^1}S_r^2S_w^2)-\bH(Y|MC_{S_r^1}S_r^2S_w^2)\\
&\qquad+\bH(M|YS_r^2S_w^2)+\bI(M;C_{S_r^1})\\
&\leq \bH(Y|C_{S_r^1}S_r^2S_w^2)-\bH(Y|MC_{S_r^1}CS_r^2S_w^2)\\
&\qquad+\bH(M|YS_r^2S_w^2)+\bI(M;C_{S_r^1})\\
&\overset{(2)}{\leq} \bH(Y|C_{S_r^1}S_r^2S_w^2)-\bH(E_2|MC_{S_r^1}CS_r^2S_w^2)\\
&\qquad+\bH(M|YS_r^2S_w^2)+\bI(M;C_{S_r^1})\\
&\overset{(3)}{=} \bH(Y|C_{S_r^1}S_r^2S_w^2)-\bH(E_2|S_r^2S_w^2)\\
&\qquad+\bH(M|YS_r^2S_w^2)+\bI(M;C_{S_r^1})\\
\end{split}
\end{equation}
(1) is from $\{S_r^2, S_w^2\}\rightarrow C_{S_r^1}\rightarrow M$ from which it follows that $\Pr (M| C_{S_r^1}S_r^2 S_w^2)=
\Pr (M| C_{S_r^1})$. The Markov chain holds because knowledge of $C_{S_r^1 } $ implies that subset pair $\{S_r^1, S_w^1\}$ is used,
 and so $\{S_r^2, S_w^2\}$  does not provide extra information; (2) is by noting that $E_2=Y-C$ if the adversary selects $\{S_r^2,S_w^2\}$. (3) is from the Markov chain 
$MC_{S_r^1}C\rightarrow \{S_r^2,S_w^2\}\rightarrow E_2$, which implies $\bI(MC_{S_r^1}C; E_2 | S_r^2S_w^2)=\bH(E_2 | S_r^2S_w^2)-\bH(E_2 | MC_{S_r^1}CS_r^2S_w^2)=0$.

So we have 
\[
\begin{split}
\bH(M)& \leq \bH(Y|C_{S_r^1}S_r^2S_w^2)-\bH(E_2|S_r^2S_w^2)+\bH(M|YS_r^2S_w^2)\\
&+\bI(M;C_{S_r^1})
\end{split}
\]
We can upper bound  $\bH(M)$ by bounding the four terms on the  right hand side of the inequality.

First, we have the bound, $\bH(Y|C_{S_r^1}S_r^2S_w^2)\leq (1-\rho_r)N\log |\Sigma|$.   
Let $[N] \setminus {S_r^1}$ be the subset of $[N]$ that is not in $S_r^1$, and $Y_{[N] \setminus {S_r^1}}$ be the components of $Y$ on the set $[N]\setminus {S_r^1}$.  Since $S_r^1\cap S_w^2=\emptyset$, if the adversary selects the set pair $\{S_r^2, S_w^2\}$, the components of $Y$ on the set $S_r^1$ will not have  error and will be equal to the components of $C$ on $S_r^1$. That is,
\begin{equation}\label{eq_up001}
\bH(Y_{S_r^1}|C_{S_r^1}S_r^2S_w^2)=0
\end{equation}
So we have, 
\[
\begin{split}
&\bH(Y|C_{S_r^1}S_r^2S_w^2)=
\bH(Y_{S_r^1}Y_{[N]\setminus {S_r^1}}|C_{S_r^1}S_r^2S_w^2)\\
&=\bH(Y_{S_r^1}|C_{S_r^1}S_r^2S_w^2)+\bH(Y_{[N]\setminus {S_r^1}}|C_{S_r^1}Y_{S_r^1}S_r^2S_w^2)\\
&\leq \bH(Y_{S_r^1}|C_{S_r^1}S_r^2S_w^2)+\bH(Y_{[N]\setminus {S_r^1}})\\
&\leq \log |Y_{[N]\setminus {S_r^1}}|\\
&\leq (1-\rho_r)N\log |\Sigma|
\end{split}
\]

To bound the second item notice that  
if the adversary selects $\{S_r^2, S_w^2\}$,  $E_2$ is uniformly distributed and so,  
\begin{equation}\label{eq_up002}
\bH(E_2|S_r^2S_w^2)=\rho_wN \log |\Sigma|
\end{equation}

From Lemma \ref{le_up4} and \ref{le_up5}, we also have the bounds $\bH(M|YS_r^2S_w^2)\leq 2\bH(\delta)+2\delta N\log |\Sigma|$ and $\bH(M)-\bH(M|C_{S_r^1})\leq 2\epsilon\rho_rN\log\frac{|\Sigma|}{\epsilon}$. So the upper bound on $\bH(M)$ is,
\[
\begin{split}
\bH(M)\leq & (1-\rho_r-\rho_w)N\log|\Sigma|+2\bH(\delta)+2\delta N\log |\Sigma|\\
&+2\epsilon\rho_rN\log\frac{|\Sigma|}{\epsilon}
\end{split}
\]
Since the message is uniformly distributed, we have $\bH(M)=\log |\cM|$.  Since for $0<\delta<\frac{1}{2}$ it holds $\delta<\bH(\delta)$, the upper bound on the rate of an AWTP code of length $N$ is obtained by usingt, $\bH(\delta)+\delta N\log |\Sigma|\leq 2\bH(\delta) N\log |\Sigma|$. That is, 
\[
\begin{split}
R(C^N)&= \frac{\log |\cM|}{N\log |\Sigma|}\\
&\leq  1-\rho_r-\rho_w+2\epsilon\rho_r(1+\log_{|\Sigma|}\frac{1}{\epsilon})+4\bH(\delta)
\end{split}
\]

\end{IEEEproof}

The following is the  proof of Theorem \ref{the_upper1}.

\begin{IEEEproof} (Theorem \ref{the_upper1})
Proof is by contradiction.  
Suppose there is a   code family $\mathbb{C}$ with achievable rate $\mathsf{C}^\epsilon 
 = 1-\rho_r-\rho_w+ 2\epsilon\rho_r (1+\log_{|\Sigma|}\frac{1}{\epsilon})+\hat{\xi} $,
 for some small constant $0< \hat{\xi}< \frac{1}{2}$.

Let $\bH(p_0)=\frac{\hat{\xi}}{8}$. For any $\hat{\xi}'\leq p_0$, we have $4\bH(\hat{\xi}')\leq \frac{\hat{\xi}}{2}$ and $\hat{\xi}'\leq \bH(\hat{\xi}')\leq \frac{\hat{\xi}}{8}$. From  Definition \ref{def_awtpfamily}, for any $0<\hat{\xi}'\leq p_0$, there is  an $N_0$ such that for any $N>N_0$,  we have  $\delta<\hat{\xi}'$ and,
\[
\begin{split}
&R(C^N)\geq \mathsf{C}^\epsilon -\hat{\xi}'\\
&= 1-\rho_r-\rho_w+2\epsilon\rho_r (1+\log_{|\Sigma|}\frac{1}{\epsilon})+\hat{\xi} -\hat{\xi}' \\
&\overset{(1)}{=} 1-\rho_r-\rho_w+2\epsilon\rho_r (1+\log_{|\Sigma|}\frac{1}{\epsilon})+ 4\bH(\delta) +\frac{\hat{\xi}}{2} -\hat{\xi}' \\
&\overset{(2)}{>} 1-\rho_r-\rho_w+2\epsilon\rho_r (1+\log_{|\Sigma|}\frac{1}{\epsilon})+ 4\bH(\delta)
\end{split}
\]
(1) is from $\bH(\delta)\leq \bH(\hat{\xi}')< \frac{\hat{\xi}}{8}$; (2) is from $\hat{\xi}'< \frac{\hat{\xi}}{2}$.

This contradicts the bound on $R(C^N)$ in Lemma \ref{le_upper1}, and so,
\[
{\mathsf C}^{\epsilon} \leq 1-\rho_r-\rho_w+2\epsilon \rho_r (1+\log_{|\Sigma|}\frac{1}{\epsilon}) 
\]

\end{IEEEproof}

For $\epsilon=0$, we have the upper bound on the  achievable rate of an AWTP code family with perfect secrecy.

\begin{corollary}\label{le_up2}
The upper bound on the achievable rate of a perfectly secure AWTP code family for a $(\rho_r, \rho_w)$-AWTP channel is,
\[
\mathsf{C}^0\leq 1-\rho_r-\rho_w
\]
\end{corollary}

\subsection{Restricted AWTP channels}
Note that the above  proof is general in the sense that the sets $S_r$ and $S_w$ can have nonempty intersection.
Restricted channels limit the adversary to the case that $S_r=S_w$.  Using the same approach, we can derive the following 
bounds on $\mathsf{C}^0$ and $\mathsf{C}^\epsilon$.

\begin{corollary}\label{le_up3} 
The upper bound of rate of {\em restricted-AWTP code family} with 
over $(\rho_r, \rho_w)$-AWTP channel is, for a perfectly secure code family,
\[
\mathsf{C}^0\leq 1-\rho_r-\rho_w
\]
and for an $\epsilon$-secure code family, 
\[
\mathsf{C}^\epsilon\leq 1-\rho_r-\rho_w+2\epsilon\rho_r (1+\log_{|\Sigma|}\frac{1}{\epsilon}) 
\]
\end{corollary}
We note that a more direct proof of Theorem \ref{the_upper1} is to use an adversary strategy 
in which $S_r\cap S_w =\emptyset$.  However this proof cannot be used for the subclass of restricted 
AWTP because, for this subclass this is not a valid adversarial strategy.  The above proof with randomized adversarial strategy
removes this restriction and  allows us to apply the same proof method
for restricted AWTP channels.


%

\section{AWTP Code Construction}\label{sec_construction}

Let $q$ to be  a prime  satisfying,  $q>Nu$. 
A message is an element of  ${\cal M}=  \mathbb{F}_q^{uRN}$, given by  ${\bf m}=\{m_1,\cdots, m_{uRN}\}\in {\cal M}$,
and
$m_i\in \mathbb{F}_q$.
 
We construct a $\del$--AWTP code  family $\mathbb{C}^0=\{C^N\}_{N\in \mathbb{N}}$,    
for a $\rorw$-AWTP channel. 
The construction uses,  (i) an FRS code, (ii) an AMD code and, (iii)  a subset evasive  sets, with the following parameters .

\begin{enumerate}
\item A $u$-Folded RS-codes of length $N$ over $\mathbb{F}_q$, with a linear algebraic decoder \cite{Gur11} using the decoding parameter $v$.

Let $\xi_1 = \xi/13$. Parameters of the FRS code are chosen as, (i) folding parameter $u=\xi_1^{-2}$, (i) decoding parameter  $v=\xi_1^{-1}$,  (iii) the length $N\geq (1/\xi_1)^{D/\xi_1 \log\log 1/\xi_1}$, and (iv) the field size satisfying $q>Nu$, and condition 2 of Theorem 3.2 in \cite{DL12}.
This latter condition on $q$ is required for efficient injective mapping into the subset evasive set, and hence efficient encoding and decoding.

\item
Assume, for simplicity, that $uR$ is an integer.  (The  argument can straighforwardly be extended to the case that this condition does not hold.)
The AMD code will be the code in Section \ref{sec_amd}, and  will have message space ${\cal X}=  \mathbb{F}_q^{uRN}$, codeword space ${\cal G} =  \mathbb{F}_q^{uRN+2}$, and $\delta\leq \frac{uR+1}{q^N}$.

\item We will use  a  $(v, v^{D\cdot  v\cdot \log\log v} )$- subspace
 evasive  sets
$\cal S$, that is  a subset of size $q^{n_1}$,  in $\mathbb{F}_q^n$, using 
the construction in Theorem 3.2 in \cite{DL12}.
 Here $D$ is a constant (See Claim 4.3 in \cite{DL12}),  and the bound  $v^{D\cdot  v\cdot \log\log v}$ on the intersection
list size of a $v$-dimensional affine subspace with $\cal S$, follows from Claim 4.3 and 3.3 in \cite{DL12}.

The parameters $n$ and $n_1$ are chosen as shown below, to achieve a rate 
$R(C^N) = 1- \rho_r-\rho_w -\xi$. 

Let $w=v^2$ and $b=\lceil\frac{uR N+2N}{w-v}\rceil$. Then we choose
$n_1=(w-v)b$, $n=wb$.

Note that $n_1$ is almost the same as the codeword length of the AMD code, i.e. $(uR+2)N$.
Also,  $v=\xi_1^{-1}$ and,
\[ n_1= \frac{w-v}{w} n= \frac{v^2-v}{v^2}n= (1-\frac{1}{v} )n= (1-\xi_1)n\]
and so the size of $\cal S$ satisfies  condition 2 in Theorem 3.2 in \cite{DL12}.

\end{enumerate}

We use $\gamma$ to denote  a primitive element of $\mathbb{F}_q$.

Let $\enc_{N}$ and
$\dec_{N}$, be the encoding and decoding algorithms of the code,   respectively. 

The constructions of the encoder and the decoder of  $C^N$ are given in Figure \ref{algo_en}.

\vspace{3mm}

\begin{center}
{\bf Figure} \ref{algo_en}
\end{center}
\begin{framed}
\noindent{\bf Encoding}: Alice does the following:

\begin{enumerate}\label{algo_en}
\item  Interpret an information block $\bf m$ of length $uRN$,   as a vector ${\bf x} \in  \bF_{q^N}^{uR}$.  
  Generate a random vector ${\bf r } \in   \bF_{q^N}$ and use it to find the codeword associated with   ${\bf x}$ 
using the AMD construction in section \ref{sec_amd},
$\mathsf{AMDenc}({\bf x})=({\bf x}, {\bf r}, {\bf t})$. The  AMD codeword is of length $uR N+2N$ over $\bF_q$.

\item   Extend  the AMD codeword to length $n_1$ by appending zeros from $\mathbb{F}_q$. Encode the AMD codeword to an element  $\bf s\in {\cal S}$, 
using the bijection mapping of the subspace evasive sets , $${\bf s}=\mathsf{SE}({\bf x}, {\bf r}, {\bf t}||0,\cdots, 0)$$  Note that elements of ${\cal S}$ are from $\bF_q^n$ and so,
$\bf s$ has length 
 $n$. 
 \item Append  a random vector ${\bf a}=(a_1 \cdots a_{u\rho_rN})\in \mathbb{F}_q^{u\rho_rN}$ 
to $\bf s$ and  form the vector that will be the message of the FRS code. Use
${\bf s}$ and ${\bf a}$ as the coefficients of the FRS codeword polynomial, $f(x)$, over $\bF_q$. 
That is $(f_0,\cdots, f_{k-1})=({\bf s}||{\bf a})$.
We have $k=deg(f)+1=  n+ u\rho_rN$. 

\item Use $\mathsf{FRSenc}$ 
to 
construct the  FRS codeword $c=\mathsf{FRSenc}(f(X))=
(c_1,\cdots ,c_N)$, with $c_i=(f(\gamma^{i(u-1)}), \cdots, f(\gamma^{iu-1})) \in \bF_q^u, \mbox{ for } i=1,\cdots, N$.
\end{enumerate}

\medskip
\noindent{\bf Decoding}: Bob does the following:
\begin{enumerate}
\item 
Let  $y=c+e$, and $w_H(e)\leq \rho_wN$.
Let,  $y= (y_1, \cdots, y_N)$ and $y_i=(y_{i, 1}, \cdots, y_{i, u})$ for $i=1, \cdots, N$. 

 Use the FRS (linear algebraic) decoding algorithm  $\mathsf{FRSdec}(y)$ to
output a matrix ${\bf M}\in \mathbb{F}_q^{k\times v}$, and a vector ${\bf z}\in \mathbb{F}_q^k$, such that the 
decoder output list 
is of the form, ${\cal L}_{\mathsf{List}} = 
 {\bf Mb}+{\bf z}$. 
${\bf M}$ has $k(= n+u\rho_rN)$ rows,  each giving a component of the output vector as a linear combination of $(b_1,\cdots, b_{v})$,
and the corresponding component of ${\bf z}$.

\item Let   $\cH$ denote  the vector space spanned by  the first $n$ equations. That is $$\cH={\bf M}_{n\times v}{\bf b}+{\bf z}_n, {\bf b}\in \mathbb{F}_q^{v},$$
where ${\bf M}_{n\times v}$ is the first $n$ rows of the submatrix of ${\bf M}$ and ${\bf z}_n$ is the first $n$ elements of $\bf z$.

 The  AWTP decoder calculates the intersection $\cS\cap \cH$ and outputs a list ${\cal L}$ of size at most $v^{D\cdot v\cdot\log\log v}$. Each codeword  is parsed and for each ${\bf s}_i\in {\cal L}$  
 a potential AMD codeword $({\bf x}_i, {\bf r}_i, {\bf t}_i)$ is generated.
For each potential AMD codeword $({\bf x}_i, {\bf r}_i, {\bf t}_i)$, the $\mathsf{AMDdec}$ checks if, ${\bf t}_i=f({\bf x}_i, {\bf r}_i)$.
 If there is a {\em  unique} valid AMD codeword in the output list of the FRS decoder, the AWTP decoder outputs the first $uRN$ components of $\bf x$ as the correct message $\bf m$.
Otherwise,   Bob  randomly selects and outputs a codeword of the AMD code.  
\end{enumerate}
\end{framed}
\vspace{2mm}

We prove secrecy and reliability of the above code, and derive the rate of the AWTP code family.

\begin{lemma}[Secrecy]\label{le_privacy}
The AWTP code $C$ above provides perfect security for $(\rho_r, \rho_w)$-AWTP channels. 
\end{lemma}

\begin{IEEEproof}
It is sufficient to show that an AWTP codeword sent over
a $(\rho_r, \rho_w)$-AWTP channel does not leak any information about the encoded element in subspace evasive sets, which includes the message sent by Alice.

The codeword polynomial is of degree $n+u\rho_r N-1 $ and so has $n+u\rho_r N$ coefficients,
  $u\rho_r N $ 
of which are randomly chosen. 
The adversary sees  $\rho_r N$ elements of $\bF_q^u$, each corresponding to a linear  equation on the 
coefficients. 
This means that 
the adversary has no information about the remaining $n$ coefficients corresponding to the message. 
The FRS coding  can be seen as coset coding in 
 \cite{OW84}, and so inheriting the security of these codes. Hence for an adversary observation $\mathsf{View}_{\cal A}=\{c_{j_1},\cdots, c_{j_{\rho_rN}}\}$,
\[
\bH(S| \mathsf{View}_{\cal A})=\bH(S)
\]
where $S$ is the element of the subspace evasive sets which is the encoding from the message $M$.
\end{IEEEproof}

\begin{lemma}[Reliability]\label{le_decodingerror1}
i) Given $N\geq v^2$, 
the AWTP code $C^N$ described above provides reliability for a $(\rho_r, \rho_w)$-AWTP channel if the following holds:
\begin{equation}\label{eq_awtppf_rate2}
\rho_w<\frac{v}{v+1}-\frac{v}{v+1}\frac{\frac{v}{v-1}(uR+3)+u\rho_r}{u-v+1}.
\end{equation}

ii) The decoding error probability of  
$\dec$ is bounded by $\delta\leq \frac{v^{D\cdot v\cdot\log\log v}}{q^N}$.
\end{lemma}



\begin{IEEEproof}
i)
FRS decoding algorithm $\mathsf{FRSdec}$ \cite{Gur11} requires,
\begin{equation}\label{eq233}
N - \rho_wN > N(\frac{1}{v + 1} + \frac{v}{v + 1}\frac{uR_{\mathsf{FRS}}}{ u - v + 1}) 
\end{equation}
The dimension of the FRS code is bounded by, 
\begin{equation}\label{eq232}
\begin{split}
k&=uR_{\mathsf{FRS}} N=u\rho_rN+n\\
&=u\rho_rN+w\lceil \frac{uR N+2N}{w-v}\rceil \\
&\overset{(1)}{\leq} u\rho_r N+\frac{w}{w-v} (uRN+3N)
\end{split}
\end{equation}
(1) is from $N \geq v^2$.  

Thus, we have,   $uR_{\mathsf{FRS}} \leq u\rho_r +\frac{w}{w-v} (uR+3)$, and replacing $R_{\mathsf{FRS}}$
in 
the decoding condition for FRS code (\ref{eq233}) gives,  
\remove{ 
\[
\rho_wN > N(\frac{1}{ v+1} + \frac{v}{v+1} \frac{\frac{w}{w-v}(uR+3)+u\rho_r}{u-v+1} )
\]
Using $w = v^2$, we have,
}
\[
\rho_w< \frac{v}{v+1} - \frac{v}{v+1}\frac{\frac{v}{v-1}(uR+3)+u\rho_r}{u-v+1}
\]

ii)
There is a decoding error  if there are at least codewords in the FRS decoder output list, that are AMD encodings of 
two messages ${\bf m}'\neq {\bf m}$.

First note that the correct message is always in the decoder list. 
This is because the FRS decoder output, 
combined with the subspace evasive sets intersection algorithm,  gives all
codewords that are at distance at most $\rho_wN$ from the received word $y$,  {\em and}, have messages that are
elements of the subset evasive set.
This list includes the  sent  codeword as the message $\bf m$ had been mapped to $\cal S$; that is the information  is
first, encoded using 
AMD coding
$({\bf x}, {\bf r}, {\bf t})=\mathsf{AMDenc}({\bf m})$,
and then mapped  to the subspace evasive sets, 
\[
\mathsf{SE}({\bf x}, {\bf r}, {\bf t}||0,\cdots, 0)\in \cL=\cS\cap \cH\;\; \mathsf{and}\;\; {\bf t}=f({\bf x}, {\bf r}),
\]
and finally,  encoded using FRS code.
Hence  the sent codeword is in the output list of the FRS decoder.

Next, we show  that
the probability that  the message associated with  any other codeword in the decoder list 
is a valid AMD codeword, is small. That is,
\[
\begin{split}
\bP( [\mathsf{SE}({\bf x}', {\bf r}', {\bf t}')\in \cS\cap \cH]\wedge [{\bf t}'=f({\bf x}', {\bf r}')])\leq \frac{uR+1}{q^N}
\end{split}
\]
From Lemma \ref{le_privacy}, the adversary has no information about the encoded subspace evasive sets element, $\bf s$, and so the AMD codeword $\mathsf{SE}({\bf x}, {\bf r}, {\bf t})={\bf s}$. This means that the adversary's error, 
$(\Delta {\bf x}_i={\bf x}'-{\bf x}, \Delta {\bf r}_i={\bf r}'-{\bf r}, \Delta {\bf t}_i={\bf t}'-{\bf t})$, is  
 independent of $({\bf x}, {\bf r}, {\bf t})$. According to Lemma \ref{le_amd}, the probability that a tampered AMD codeword, $({\bf x}', {\bf r}', {\bf t}')$, passes the verification is no more than $\frac{uR+1}{q^N}$. 

Finally, we show the probability of decoding error is 
at most $\delta \leq \frac{v^{D\cdot v\cdot \log\log v}}{q^N}$. The list size is at most $|\cS\cap \cH|\leq v^{D'\cdot v\cdot \log\log v}$ and, $uR+1 \leq u+1=v^2+1$. 
So by using the union bound and letting  $D=D'+3$,   the probability that some $({\bf x}', {\bf r}', {\bf t}')\neq ({\bf x}, {\bf r}, {\bf t})$ in the decoded list passes the verification ${\bf t}'=f({\bf x}', {\bf r}')$, is no more than $\frac{v^{D\cdot v\log\log v}}{q^N}$. 
 \remove{
 That is
\[
\begin{split}
&\bP(\bigcup_{ {\bf s}'\in \cL}({\bf x}', {\bf r}', {\bf t}')=\mathsf{SE}^{-1}({\bf s}')\wedge {\bf t}'=f({\bf x}', {\bf r}'))\\
&\leq \sum_{ {\bf s}'\in \cL}\bP(({\bf x}', {\bf r}', {\bf t}')=\mathsf{SE}^{-1}({\bf s}')\wedge {\bf t}'=f({\bf x}', {\bf r}'))\\
&\leq \sum_{ {\bf s}'\in \cL}\bP({\bf t}'=f({\bf x}', {\bf r}'))
\leq \frac{\ell|\cL|}{q^N}
\leq \frac{v^{(D+2)\cdot v\cdot \log\log v}}{q^N}
\end{split}
\]
}
\end{IEEEproof}

\subsection{Secrecy Capacity}
The achievable rate of the code family $\mathbb{C}= \{C^N \}_{N\in \mathbb{N}}$ is given by the following Lemma.

\remove{
\begin{lemma} \label{le_informationrate1}
The AWTP code $C^N$ described above provides reliability for a $(\rho_r, \rho_w)$-AWTP channel if the following holds:
\begin{equation}\label{eq_awtppf_rate2}
\rho_w<\frac{v}{v+1}-\frac{v}{v+1}\frac{\frac{v}{v-1}(uR+3)+u\rho_r}{u-v+1}.
\end{equation}
\end{lemma}

\begin{IEEEproof}
FRS decoding algorithm $\mathsf{FRSdec}$ \cite{Gur11} requires,
\begin{equation}\label{eq233}
N - \rho_wN > N(\frac{1}{v + 1} + \frac{v}{v + 1}\frac{uR_{\mathsf{FRS}}}{ u - v + 1}) 
\end{equation}
The dimension of the FRS code is bounded by,
\begin{equation}\label{eq232}
\begin{split}
k&=uR_{\mathsf{FRS}} N=u\rho_rN+n
=u\rho_rN+w\lceil \frac{\ell N+2N}{w-v}\rceil \\
&\overset{(1)}{\leq} u\rho_r N+\frac{w}{w-v} (uRN+3N)
\end{split}
\end{equation}
Inequality (1) holds because $\ell \leq uR+1$.

Replacing $R_{\mathsf{FRS}}$ by (\ref{eq232}), the decoding condition for FRS code (\ref{eq233}) holds if, 
\[
\rho_wN > N(\frac{1}{ v+1} + \frac{v}{v+1} \frac{\frac{w}{w-v}(uR+3)+u\rho_r}{u-v+1} )
\]
Using $w = v^2$, we have,
\[
\rho_w< \frac{v}{v+1} - \frac{v}{v+1}\frac{\frac{v}{v-1}(uR+3)+u\rho_r}{u-v+1}
\]
\end{IEEEproof}


}
\begin{lemma}[Achievable Rate of  $\mathbb{C}$]\label{the_informationrate2}
The information rate of the  AWTP code family $\mathbb{C}=\{C^{N}\}_{N\in \mathbb{N}}$ for a $(\rho_r, \rho_w)$-AWTP channel is $R(\mathbb{C})=1-\rho_r-\rho_w$. 
\end{lemma}
\begin{IEEEproof}
For a given  small  $0< \xi < \frac{1}{2}$, let the code parameters be chosen as, $\xi_1=\frac{\xi}{13}$,
 $v=1/\xi_1$ and $u=1/\xi_1^2$. 
Finally let, $N_0>(1/\xi_1)^{D/\xi_1\log\log 1/\xi_1}$ where $D>0$ is a constant.

We have, starting from the right hand side of (\ref{eq_awtppf_rate2}), 

\begin{eqnarray*}
&&\frac{v}{v+1} - \frac{v}{v+1}\frac{\frac{v}{v-1}(uR+3)+u\rho_r}{u-v+1}\\
&&\qquad\qquad=\frac{1}{\xi_1+1}-\frac{1}{\xi_1+1}\frac{\frac{1}{1-\xi_1}(R+3\xi_1^2)+\rho_r}{1-\xi_1+\xi_1^2} \\
&&\qquad\qquad\overset{(1)}\geq \frac{1}{\xi_1+1}- \frac{\frac{1}{1-\xi_1}(R+3\xi_1^2)+\rho_r}{1+\xi_1^3}\\
&&\qquad\qquad\overset{(2)}\geq 1-\xi_1-(\frac{1}{1-\xi_1}(R+3\xi_1^2)+\rho_r)\\
&&\qquad\qquad\overset{(3)}\geq 1-\xi_1-((1+2\xi_1)(R+3\xi_1)+\rho_r)\\
&&\qquad\qquad= 1-\xi_1-(R+11\xi_1+\rho_r)\\
&&\qquad\qquad=1- R-\rho_r-12\xi_1
\end{eqnarray*}

Here (1) is by multiplying the numerator and denominator of the first term on the LHS by $1-\xi_1$ and then ignoring $\xi_1^2$ and 
$\xi_1^3$ in the denominators of the two terms, (2) is by multiplying the numerator and denominator of the first term in the outer parenthesis by $1+2\xi_1$, and then ignoring $\xi_1$ in the denominator, and (3) is from $\frac{1}{1-\xi_1}\leq 1+2\xi_1$ since $\xi_1\leq \frac{1}{2}$.
\remove{
\[
\begin{split}
&1- R-\rho_r-12\xi_1
=1-\xi_1-(R+11\xi_1+\rho_r)\\
&< 1-\xi_1-((1+2\xi_1)(R+3\xi_1)+\rho_r)\\
&\leq 1-\xi_1-(\frac{1}{1-\xi_1}(R+3\xi_1^2)+\rho_r)\\
&\leq \frac{1}{\xi_1+1}- \frac{\frac{1}{1-\xi_1}(R+3\xi_1^2)+\rho_r}{1+\xi_1^3}\\
&= \frac{1}{\xi_1+1}-\frac{1}{\xi_1+1}\frac{\frac{1}{1-\xi_1}(R+3\xi_1^2)+\rho_r}{1-\xi_1+\xi_1^2}\\
&=\frac{v}{v+1} - \frac{v}{v+1}\frac{\frac{v}{v-1}(uR+3)+u\rho_r}{u-v+1}
\end{split}
\]
}
Hence the decoding condition (\ref{eq_awtppf_rate2}) of AWTP code
 is  satisfied for, 
\begin{equation}\label{pf_awtp_rate1}
\rho_w= 1- R-\rho_r-12\xi_1.
\end{equation}
and so,
$
R(C^N)=1-\rho_r -\rho_w -12\xi_1
$.

Now since $\xi= 13\xi_1$, for any $N>N_0$, the rate of the AWTP code $C^N$ is 
\[
\begin{split}
\frac{1}{N}\log_{|\Sigma|}|{\cal M}|& = R(C^N) = 1-\rho_r-\rho_w-12\xi_1\\
&>1-\rho_r-\rho_w-\xi=R(\mathbb{C})-\xi
\end{split}
\]
and the probability of the decoding error is bounded as,
\[
\delta\leq (1/\xi_1)^{D/\xi_1\log\log 1/\xi_1}q^{-N}\leq Nq^{-N} \leq \xi
\]
This concludes that the achievable rate of AWTP code family $C$ is $R(\mathbb{C})=1-\rho_r-\rho_w$.

\end{IEEEproof}

The   computational complexity  of encoding is $\mathcal{O}((N\log q)^2)$. The  combined computational complexity of the FRS decoding algorithm and  subspace evasive sets intersection algorithm is,  $\mathsf{Poly}((1/\xi)^{D/\xi\log\log 1/\xi})$. An AMD verification costs  $\mathcal{O}((N\log q)^2)$, and so the total  complexity  of the AWTP decoding is $\mathsf{Poly}(N)$.

\begin{theorem}\label{the_awtpcode}
For any small $\xi > 0$, there is a $(0,\delta)$-AWTP code $C^N$ of length $N$ for $(\rho_r,\rho_w)$-AWTP
channel, such that the information rate is $R(C^N ) = 1-\rho_r-\rho_w-\xi$, the alphabet size is $|\Sigma| = \mathcal{O}(q^{1/\xi^2})$,
and the decoding error is bounded as, $\delta < q^{-\mathcal{O}(N)}$. The decoding computation is $\mathsf{Poly}(N)$. The AWTP code family $\mathbb{C}=\{C^{N}\}_{N\in \mathbb{N}}$ achieves secrecy capacity ${\mathsf C}^0= R(\mathbb{C})=1-\rho_r-\rho_w$ for a  $(\rho_r, \rho_w)$-AWTP channel.
\end{theorem}


\section{AWTP Codes and 1-round SMT}\label{sec_smtawtp}

\subsection{Secure Message Transmission (SMT)}

A  cryptographic primitive that is closely  related to AWTP model is SMT with a number of variations 
in definition and setting. 
We only consider  {\em 1-round $(\epsilon, \delta)$-SMT protocols} and a definition of reliability first considered in \cite{FW00}.

 In a 1-round $(\epsilon, \delta)$-SMT protocol a  sender and a receiver in a network, 
 are connected by $N$ vertex-disjoint paths, referred to as {\em wires}. 
 The goal is to enable sender to send a message $m\in \mathcal{M}$ with a probability distribution $\bP(M)$,
 to the receiver such that the  receiver receives \emph{reliably} and \emph{privately}, where $\delta$ and
 $\epsilon$ bound the error probability and loss of information, respectively, as defined below.
In SMT setting the adversary $\cal A$ has unlimited computational power and can corrupt and control a subset of wires: the adversary can eavesdrop, block or modify
what is sent over  the  corrupted wires.  We consider an adaptive {\em threshold adversary} that can corrupt at most $t$ out of the $N$ wires. 

A  {\em 1-round $(\epsilon, \delta)$-SMT protocol} has two algorithms. A probabilistic {\em encoding algorithm $\mathsf{SMTenc}$ }and a deterministic {\em decoding algorithm $\mathsf{SMTdec}$.}
Let  $\mathsf{View}_{\cal{A}}(\mathsf{SMTenc}(m), r_{\cal A})$ denote  the  adversary's view of the communication over the $t$ wires 
that have been corrupted, when message $m$ has been sent and the adversary's randomness has been $r_{\cal A}$. 
\begin{definition}\label{def_SMT}
An ($\epsilon_{\mathsf{SMT}}, \delta_{\mathsf{SMT}})$-Secure Message Transmission (
SMT) protocol satisfies  the following two properties:
\begin{itemize}
\item {Secrecy:} For any two messages $m_1, m_2 \in \mathcal{M}$,

\[
\begin{split}
&\ADV^{\mathsf{ds}}(\mathsf{SMTenc}, \mathsf{View}_{\cal A})\stackrel{\triangle}=\\
&\qquad \max_{m_1,m_2}\SD(\mathsf{View}_{\cal{A}}(\mathsf{SMTenc}(m_1), r_{\cal A}) \\
&\qquad\qquad\qquad\;\; \mathsf{View}_{\cal{A}}(\mathsf{SMTenc}(m_2), r_{\cal A}))\leq \epsilon_{\mathsf{SMT}}
\end{split}
\]
	
\item {Reliability:} Receiver $\cal R$ outputs the wrong message with probability no more than $\delta_{\mathsf{SMT}}$. 
\[
\bP(M_\cS\neq M_\cR ) \leq \delta_{\mathsf{SMT}}
\]
\end{itemize}
\end{definition}
\vspace{-2mm}
When it is clear  from the context, we omit the subscript ``${\mathsf{SMT}}$" and simply use
  $(\epsilon,\delta)$-SMT. 

 \remove{
 This definition of reliability for SMT was first proposed in \cite{KS09} 
and  compared to the conventional definition of reliability \cite{FW00} that allows the receiver to output incorrect messages, has the advantage of
guaranteed correctness for 
 the decoder's output message.
 }
A   {\em perfect  SMT protocol } has   $\epsilon_{\mathsf{SMT}} = 0$ and $\delta_{\mathsf{SMT}} = 0$.
It was proved \cite{FW00} that $(\epsilon_{\mathsf{SMT}},\delta_{\mathsf{SMT}})$-SMT for $\delta_{\mathsf{SMT}} <\frac{1}{2}(1-\frac{1}{|\cM|})$, is possible only if $N\geq 2t+1$  and 1-round $(0,0)$-SMT is possible only if $N\geq 3t+1$ \cite{DDWY93}.  Let ${\cal V}_i$ denote the set of possible transmissions (also called {\em transcripts}) of each wire. 
 \emph{Transmission rate of an SMT protocol} is defined
$\tau_R(\mathsf{SMT})=\frac{\mathsf{Total \ Length\ of\ transcript}}{\mathsf{Length\ of\ message}}=\frac{\sum_i \log |{\cal V}_1|}{\log |\cM|}$.

For 1-round $(0,0)$-SMT protocols,  the lower bound on  transmission rate is  $\frac{N}{N-3t}$  \cite{FFGV07}, 
and for (0, $\delta_{\mathsf{SMT}}$)-SMT, the bound  is $\frac{N}{N-2t}$ \cite{PCSR10}.
1-round $(0,0)$-SMT and  
  (0, $\delta_{\mathsf{SMT}}$)-SMT protocols
 whose transmission rates asymptotically reach $\mathcal{O}(\frac{N}{N-3t})$ and $\mathcal{O}(\frac{N}{N-2t})$, respectively, 
are called transmission {\em optimal. }

\subsection{Relation between AWTP Code and SMT}

\edwtp\;codes are closely related to  1-round $(\epsilon_{\mathsf{SMT}}, \delta_{\mathsf{SMT}})$-SMT protocols. In the following we show the relationship between the two primitives. 

\begin{definition}[Symmetric SMT]
An SMT protocol is called symmetric SMT if the protocol remains invariant under any permutation of  the wires.
\end{definition}
Let $(W^i_1, W^i_2\cdots W^i_N)$  denote the set of possible transmissions on the $N$ wires in an $r$-round SMT protocol.
In a symmetric protocol, for each round $i$, we have $W^i_j= W, j=1\cdots N$. That is the set of possible transmissions on a wire,
is independent of the wire. All known constructions of SMT protocols are symmetric.

\begin{theorem} \label{le_c1}
There is a one-to-one correspondence between 
 an \edwtp\;code $C^N$ of length $N$ that provides security for a restricted AWTP channel with $S=S_r=S_w$, 
and a   1-round $(\epsilon_{\mathsf{SMT}},\delta_{\mathsf{SMT}})$ symmetric SMT protocol for $N$ wires with security against a $(t,N) $ threshold adversary. \\
Furthermore, an \edwtp\;code for a $(\rho_r, \rho_w)$-AWTP channel 
can be used to  construct a code for a restricted AWTP channel, resulting in a 
 1-round $(\epsilon_{\mathsf{SMT}},\delta_{\mathsf{SMT}})$ symmetric SMT.  
\end{theorem}
\begin{IEEEproof}
 Consider an $(\epsilon,\delta)$-AWTP code $C^N$ over a restricted AWTP channel with $S=S_r=S_w$. By associating each component of the code with a distinct wire, one can construct a 1-round $(\epsilon_{\mathsf{SMT}},\delta_{\mathsf{SMT}})$ symmetric SMT protocol for $ N$ wires. The protocol
 security   is against a threshold $(t,N)$ adversary
with $t= \rho N$.
The SMT encoding and decoding are obtained from the corresponding functions in the \edwtp\; code;  that is,   $\mathsf{SMTenc}(m, r_{\cal S})= \enc(m, r_{\cal S})$ and  $\mathsf{SMTdec}(y)= \dec(y)$. 
To relate the security and reliability of the SMT protocol to those of the AWTP-code, we note the following:
\begin{enumerate}
\item  Definitions of privacy in the two primitives are both in terms of the
statistical distance of the adversary's view for two messages chosen by the adversary (Compare definition \ref{def_SMT} and definition \ref{def_awtpcode}). 
\item Definitions of error in decoding for the two primitives both requires the decoder to output the correct message with probability at least $1-\delta$. 

\item  Adversary's capabilities in the two models are the same.  The corruption of the codeword in  \edwtp\; code is by an additive error, while in SMT  the adversary can  arbitrarily modify the   $|S|=t$ wires. 
However for restricted AWTP channels with  $S=S_r=S_w$, modifying $t$ components $(c_{i_1}, \cdots c_{i_t})$ to $(c'_{i_1}, \cdots c'_{i_t})$ is equivalent to ``adding" the error $e$ with $\mathsf{SUPP}(e) =S$ and $(e_{i_1}, \cdots e_{i_t}) = ((c'_{i_1}-c_{i_1}), \cdots (c'_{i_t} -c_{i_t}))$ and so for these channels additive errors cover all possible adversarial 
tampering.
\end{enumerate}

The theorem follows by  constructing a restricted \edwtp\; code with $S=S_r=S_w$ from a   1-round $(\epsilon_{\mathsf{SMT}}, \delta_{\mathsf{SMT}})$ symmetric SMT,  using the same association of the code components and the wires. We will have $\epsilon=\epsilon_{\mathsf{SMT}}$ and
$ \delta=\delta_{\mathsf{SMT}}$.

\end{IEEEproof}

Corollary \ref{c2} below follows from the one-to-one correspondence established  in Theorem \ref{le_c1}.
\remove{between the two primitives and the definitions of  the transmission rate of 1-round $(\epsilon_{\mathsf{SMT}}, \delta_{\mathsf{SMT}})$-SMT protocols w
and the  rate 
 of $(\rho_r, \rho_w)$-AWTP codes for restricted AWTP channels, we have the following.
}

\begin{corollary} \label{c2}
Let $R(C^N)$ be the rate of an $(\epsilon, \delta)$-AWTP code $C^N$ for a restricted AWTP channel. 
The transmission rate of 
the associated 1-round $(\epsilon_{\mathsf{SMT}}, \delta_{\mathsf{SMT}})$ symmetric SMT is given by, 
$\tau_R({\mathsf{SMT}})=\frac{N\log {|\cal V}|}{\log |{\cal M}|} = \frac{1}{R(C^N)}$. 
\end{corollary}

The upper bound on the secrecy rate  (Lemma \ref{le_up3}) of
$(0, \delta)$-AWTP codes for restricted AWTP channels, 
gives a lower bound on the transmission rate of  1-round 
  $(0, \delta_{\mathsf{SMT}})$ symmetric SMT protocols.
\begin{theorem}\label{the_smtbound}
For a 1-round $(\epsilon_{\mathsf{SMT}}, \delta_{\mathsf{SMT}})$ symmetric  SMT protocol, transmission rate is lower bounded as,
\[
\tau_R(\mathsf{SMT})\geq\frac{N}{N-2t+2t\epsilon (1+\log_{|\cal V|}\frac{1}{\epsilon})}.\]
For $\epsilon_{\mathsf{SMT}}=0$, 
the bounded reduces to the known bound,
 $\tau_R(\mathsf{SMT})\geq\frac{N}{N-2t}$  \cite{PCSR10}.
\end{theorem}
\begin{IEEEproof}
Using Theorem \ref{le_up3}, for a   1-round $(\epsilon, \delta)$ symmetric  SMT over $N$ wires with $t=\rho N$, 
there is a corresponding \edwtp\; code for a restricted  AWTP channel with $S=S_r=S_w$ whose 
information rate  is upper bounded by,
\[
R(C^N)\leq 1-2\rho+2\epsilon \rho(1+\log_{|\Sigma|}\frac{1}{\epsilon})\]
Since the transmission rate of an  $(\epsilon, \delta)$ symmetric SMT protocol is the inverse of the
information rate of the corresponding \edwtp\; code, we have 
\[
\begin{split}
\tau_R({\mathsf{SMT}}) &= \frac{1}{R(C^N)}\\
&\geq \frac{1}{1-2\rho+2\epsilon \rho(1+\log_{|\cal V|}\frac{1}{\epsilon})}\\
&= \frac{N}{N-2t+2t\epsilon (1+\log_{|\cal V|}\frac{1}{\epsilon})}\\
\end{split}
\]
\end{IEEEproof}
It has been proved \cite{FW00}  that for 1-round $(\epsilon,\delta)$-SMT protocols for $\delta \leq \frac{1}{2}(1-\frac{1}{|\cM|})$  can be constructed only if, $N\geq 2t+1$.
\begin{corollary}
For $N=2t+1$, we will have,   
\[
\tau_R({\mathsf{SMT}}) = \frac{1}{R(C^N)}\geq  \frac{2t+1}{1+2t\epsilon (1+\log_{|\cal V|}\frac{1}{\epsilon})}
\]
\end{corollary}
This is the first and the only known lower bound on the  transmission rate of $(\epsilon, \delta)$ symmetric SMT protocols.
Using a similar approach one can obtain an alternative proof for the 
known  lower bound on the transmission rate of 1-round $(0,  \delta)$ symmetric  SMT protocols (Theorem 10,  \cite{PCSR10})


\section{Concluding Remarks}

We proposed a  model for active adversaries in wiretap channels, derived secrecy capacity and
gave an explicit construction for a family of capacity achieving codes.
The model is a natural extension of Wyner wiretap models when the adversary is 
a powerful active adversary that uses its partial observation of the 
communication channel to introduce adversarial noise in the channel.
The adversary's view of communication is the same as wiretap II model.
However unlike noiseless main channel in wiretap II,  we allow the main channel 
to be  corrupted by the  adversary inline with the corruption 
introduced by the adversary in Hamming's model of reliable communication.
This is the first model of adversarial wiretap where
the adversary's view is used by the adversary for forming its adversarial noise.
All previous work (See Section \ref{sec_relatedwork}) assume the view of the the eavesdropper 
does not affect the noise added to the channel.
\\
AWTP model  provides a
 framework
 for studying SMT 
which so far  has been studied independent of wiretap model.
The fruitfulness of this  was demonstrated by deriving a new lower bound
on the transmission rate of 1-round $(\epsilon,\delta)$ symmetric SMT protocols. 
It is an interesting question if this bound applies to all SMT protocols. That is if allowing different 
transcript set would increase the bound. 
 \remove{Using the one-to-one correspondence between \edwtp\; codes and \ed-SMT established in 
Theorem \ref{le_c1}, one can define the  information rate of 1-round \ed-SMT protocols.

and for a family of protocols with $t = \rho N$, obtain secrecy capacity.
It is worth noting that these definitions require $t = \rho N$ and so will not appl

define a new efficiency measure for \ed-SMT protocols in terms of information rate of the
corresponding \edwtp\; code.  An \ed-SMT protocol family (defined for $N\in \mathbb{N}$) is {\em information-rate optimal}
if the information rate of the corresponding \edwtp\; code is of the order of 
$O( \frac{N}{N-2t+2t\epsilon (1+\log_{|\cal V|}\frac{1}{\epsilon})})$. We have not defined SMT family

Construction of  \ed-SMT protocol family that are  information-rate optimal is an interesting open question.
}
\\
In the discussion part of Section 1, 
we listed a number of open questions that will  improve our results.
More general settings such as allowing interaction between the sender and the receiver, 
as well as  variations in the 
adversarial power 
including considering adversarial and probabilistic noise both,  will be 
 interesting 
directions
for future work.
Another important direction  for future work is the study of key agreement problem
over an AWTP channel.


\appendices

\section{Subspace Evasive Sets}\label{ap_ses}

\subsection{Encoding Algorithm}\label{ap_sesencoding}

We show the encoding map $\mathsf{SE}: {\bf v}\rightarrow {\bf s}$. 
Assuming there is a vector $\bf v$ of length $n_1$ and $(w-v)|n_1$. First we divide the vector into $\frac{n_1}{w-v}$ blocks. Then for each block ${\bf v}_i$ for $i=1,\cdots, \frac{n_1}{w-v}$, we encode into a block ${\bf s}_i$ using bijection $\varphi$. Then we concatenate each block ${\bf s}_i$ for $i=1,\cdots, \frac{n_1}{w-v}$ and generate $\bf s$ in $\cS$. We give the function $\varphi$ in the following.

\begin{lemma}(Claim 4.1)
Assume that at least $v$ of the degree $d_1, \cdots, d_v$ are co-prime to $|\bF|-1$. Then there is an easy to compute bijection $\varphi: \bF^{w-v}\rightarrow {\bf V}_{\bF}\subset \bF^w$. Moreover, there are $w-v$ coordinates in the output of $\varphi$ that can be obtained from the identity mapping $\mathsf{Id}: \bF^{w-v}\rightarrow \bF^{w-v}$. 
\end{lemma}

Let $d_{j_1},\cdots, d_{j_v}$ be the degree among $d_1, \cdots, d_w$ co-prime to $|\bF|-1$ and let $J=\{j_1,\cdots,j_v\}$ and $x_{j_i}^{d_{j_i}}=y_{i}$. On the positions $[w]\backslash J$, the map $\varphi$ takes the elements from $\bF^{w-v}$ to $\bF^{[w]\backslash J}$. For the elements on $J$, there is
\[
\sum_{j\in J}A_{i,j}x_j^{d_j}=-\sum_{j\notin J}A_{i,j}x_j^{d_j}
\]
Let $A'$ be the $v\times v$ minor of $A$ given by restricting $A$ to columns in $J$ and $b_i=-\sum_{j\notin J}A_{i,j}x_j^{d_j}$. Then 
\[
A'y=b
\]
and for each $y$, there is unique solution of $x_{j_i}^{d_{j_i}}=y_{i}\mod q$ because $d_{j_i}$ is co-prime to $q-1$.

The computational complexity of mapping each vector ${\bf v}_i$ into ${\bf s}_i$ is $\pl(v)$. Since $\bf v$ is consisted by $b=\frac{n}{w}$ vectors, the total computational complexity of encoding a vector into an element in subspace evasive sets is $\pl(n)$.

\subsection{Intersection Algorithm}\label{ap_sesdecoding}

We show how to compute the intersection $\cS \cap \cH$ given $(v, \ell)$ subspace evasive sets $\cS$ and $v$-dimension subspace $\cH$. The subspace evasive sets $\cS$ will filter out the elements in $\cH$ and output a set of elements $\cS\cap \cH$ with size no more than $\ell$.

Because $\cH$ is $v$-dimensional subspace and $\cH\subset \bF^n$, there exists a set of affine maps $\{\ell_1, \cdots, \ell_n\}$ such that for any elements ${\bf x}=\{x_1, \cdots, x_m\}\in \cH$, there is $x_i=\ell_i(s_1, \cdots, s_v)$.

We show the result by induction of the number of blocks $i=1,\cdots, n/w$. If $i=1$, let $\cH_1:=\{(x_1, \cdots, x_w): (x_1, \cdots, x_n)\in \cH\}$, the dimension of $\cH_1$ is $r_1\leq v$ and $\cH_{x_1, \cdots, x_w}=\{(x_1, \cdots, x_n)\in \cH: (x_1, \cdots, x_w)\}$ such that $\cH=\cup_{(x_1, \cdots, x_w)\in \cH_1}\cH_{x_1, \cdots, x_w}$, and the dimension of $\cH_{x_1, \cdots, x_w}$ is $v-r_1$.
There is 
\[
\begin{split}
&{\bf V}_{\bF}(f_1,\cdots, f_v)\cap \cH_1\\
&=\{(x_1, \cdots, x_w)=(\ell_1(s_1, \cdots, s_v), \cdots, \ell_w(s_1, \cdots, s_v)):\\ 
&\qquad f_1(\ell_1(s_1, \cdots, s_v), \cdots, \ell_w(s_1, \cdots, s_v))=0,\cdots, \\
&\qquad f_v(\ell_1(s_1, \cdots, s_v), \cdots, \ell_w(s_1, \cdots, s_v))=0\}
\end{split}
\]
We can solve the $v$ equations to get $(s_1, \cdots, s_v)$ and then obtain $(x_1, \cdots, x_w)$.
Since $\cH_1\subset \bF^w$,
\[
{\bf V}_{\bF}(f_1,\cdots, f_v)\cap \cH_1={\bf V}(f_1,\cdots, f_v)\cap \cH_1
\] 
By Bezout's theorem, there is $|{\bf V}(f_1, \cdots, f_v)\cap \cH_1|\leq (d_1)^{r_1}$. So there are at most $(d_1)^{r_1}$ solutions for $(x_1, \cdots, x_w)\in \cH_1$. The computational time of solving the equation system follows from powerful algorithms that can solve a system of polynomial equations (over finite fields) in time polynomial in the size of the output, provided that the number of solutions is finite in the algebraic closure (i.e the ‘zero-dimensional’ case). So for $i=1$, there are $(d_1)^{r_1}$ solutions for $(x_1, \cdots, x_w)$. The computational time is at most $\pl((d_1)^{r_1})$.

For every fixed of the first $w$ coordinates, we reduce the dimension of $\cH$ and obtained a new subspace on the remaining coordinates. By induction, we have $|{\bf V}(f_1, \cdots, f_v)\cap \cH_{x_1, \cdots, x_w}|\leq (d_1)^{r-r_1}$ for all  $(x_1, \cdots, x_w)\in \cH_1$. Hence there is $|{\bf V}(f_1, \cdots, f_v)\cap \cH | \leq (d_1)^r$.

Similarly, we can compute all the solutions in times $\pl((d_1)^{r_1})\cdot \pl((d_1)^{r_2}) \cdots \pl((d_{1})^{r_{n/w}})$, where $r_1+r_2+\cdots +r_{n/w}=v$. So the running time of decoding algorithm is $\pl((d_1)^{v})$. Since $d_1$ can be bounded by $d_1\leq v^{D\log\log v}$ (Claim 4.3 \cite{DL12}) with constant $D$, the total running time for the intersection algorithm is $\pl(v^{D\cdot v\log\log v})$.


\section{List Decodable Code}

\subsection{Decoding algorithm of FRS code} \label{decode_FRS}

Linear algebraic list decoding \cite{Gur11} has two main steps: interpolation and message finding as outlined below.

\begin{itemize}
\item Find a polynomial, $Q(X, Y_1, \cdots, Y_v)=A_0(X)+A_1(X)Y_1+\cdots+A_v(X)Y_v$, over $\bF_q$ such that $\mbox{deg}(A_i(X)) \leq D$, for $i=1 \cdots v$, and $\mbox{deg}(A_0(X)) \leq D+k-1$, 
  satisfying $Q(\alpha_i, y_{i_1}, y_{i_2},\cdots ,y_{i_v})=0$ for $1\leq i\leq n_0$, where $n_0=(u-v+1)N$.

\item Find all polynomials $f(X) \in \bF_q[X]$ of degree at most $k-1$, with  coefficients $f_0, f_1 \cdots f_{k-1}$,
    that satisfy, $A_0(X)+A_1(X)f(X)+A_2(X)f(\gamma X)+\cdots+A_v(X)f(\gamma^{v-1}X)=0$,
by solving linear equation system.
\end{itemize}

The two above requirements are satisfied if $f \in \bF_q[X]$ is a polynomial of degree at most $k- 1$ whose FRS encoding  agrees with the received word $\bf y$ in at least $t$ components: 
\[
t>N(\frac{1}{v+1}+\frac{v}{v+1}\frac{uR}{u-v+1})
\]

This means we need to find all polynomials $f(X) \in \bF_q[X]$ of degree at most $k-1$, with coefficients $f_0, f_1, \cdots, f_{k-1}$, that satisfy, 
\[
\begin{split}
&A_0(X)+A_1(X)f(X)+A_2(X)f(\gamma X)+\cdots\\
&+A_v(X)f(\gamma^{v-1}X)=0
\end{split}
\]

Let us denote $A_i(X) = \sum_{j=0}^{D+k-1} a_{i,j}X^j$
for $0 \leq i \leq v$. ($a_{i,j} = 0$ when $i \geq 1$ and $j \geq D$). 
Define the polynomials,
\[
\begin{cases}
\begin{split}
&B_0(X)  =  a_{1,0} + a_{2,0}X + a_{3,0}X^2+ \cdots + a_{v,0}X^{v-1}\\
& \ \ \ \ \ \ \vdots\\
&B_{k-1}(X) = a_{1,k-1} + a_{2,k-1}X + a_{3,k-1}X^2+ \cdots \\
&+  a_{v,k-1}X^{v-1}\\
\end{split}
\end{cases}
\]

We examine the condition that the coefficients of $X^i$ of the polynomial $Q(X) = A_0(X) + A_1(X)f(X) + A_2(X)f(\gamma X) + \cdots + A_v(X)f(\gamma^{v-1} X)=0$ equals $0$, for $i=0\cdots k-1$. This is equivalent to the following system of linear equations for $f_0\cdots f_{k-1}$.

\begin{equation}\label{eq_FRS2}
\begin{split}
&\begin{bmatrix}
B_0(\gamma^0) &0  & 0 &\cdots  &0 \\
B_1(\gamma^0) &B_0(\gamma^1)  & 0 &\cdots  &0 \\
B_2(\gamma^0) &B_1(\gamma^1)  & B_0(\gamma^2) &\cdots  &0 \\
\vdots &\vdots  & \vdots & \vdots & \vdots \\
B_{k-1}(\gamma^0) & B_{k-2}(\gamma^{1}) &B_{k-3}(\gamma^2)  &\cdots  & B_0(\gamma^{k-1})
\end{bmatrix}\\
&\times\begin{bmatrix}
f_0\\
f_1\\
f_2\\
\vdots\\
f_{k-1}
\end{bmatrix}=\begin{bmatrix}
-a_{0,0}\\
-a_{0,1}\\
-a_{0,2}\\
\vdots\\
-a_{0,k-1}
\end{bmatrix}
\end{split}
\end{equation}
The rank of the matrix of (Eqs. \ref{eq_FRS2}) is at least $k-v+1$ because there are at most $v-1$ solutions of equation $B_0(X)=0$ so at most $v-1$ of $\gamma^i$ that makes $B_0(\gamma^i)=0$. The dimension of solution space is at most $v-1$ because the rank of matrix of (Eqs. \ref{eq_FRS2}) is at least $k-v+1$. So there are at most $q^{v-1}$ solutions to (Eqs. \ref{eq_FRS2}) and this determines the size of the list which is equal to $q^{v-1}$.


\ifCLASSOPTIONcaptionsoff
  \newpage
\fi

\end{document}